%% file: main.tex
\PassOptionsToPackage{table}{xcolor}
\AddToHook{package/hyperxmp/before}{%
  \RequirePackage[bookmarksnumbered,unicode]{hyperref}}
\documentclass[manuscript]{acmart}
\acmJournal{TOIS}
\usepackage{booktabs,makecell,multirow,tabularx,adjustbox}
\usepackage[section]{placeins}
\usepackage{graphicx} 
\usepackage{float}
\usepackage[most]{tcolorbox}
\usepackage{enumitem}
\usepackage{appendix}
\usepackage{needspace}

\definecolor{PromptTitleBg}{HTML}{4A4A4A}   
\definecolor{PromptBodyBg}{HTML}{F5F5F5}    
\definecolor{PromptFrame}{HTML}{333333}     
\definecolor{PromptAccent}{HTML}{C65D00}    

\newtcolorbox{promptbox}[1][]{
  enhanced,
  breakable,
  colback=PromptBodyBg,
  colframe=PromptFrame,
  coltitle=white,
  fonttitle=\bfseries\small,
  title filled,
  attach boxed title to top left={yshift=-2mm, xshift=4mm},
  boxed title style={
    colback=PromptTitleBg,
    arc=2mm,
    boxrule=0pt,
    left=3mm, right=3mm, top=1.2mm, bottom=1.2mm
  },
  arc=3mm,
  boxrule=0.6pt,
  left=4mm, right=4mm, top=3mm, bottom=3mm,
  before skip=8pt,
  after skip=10pt,
  #1
}

\usepackage{fvextra}
\tcbuselibrary{breakable}
\usepackage{amsthm}

\definecolor{theoremblue}{RGB}{230, 240, 250}
\definecolor{theoremborder}{RGB}{70, 130, 180}
\definecolor{remarkgray}{RGB}{245, 245, 245}
\definecolor{remarkborder}{RGB}{150, 150, 150}

\newtcbtheorem[number within=section]{mythm}{Theorem}{
    colback=theoremblue,
    colframe=theoremborder,
    fonttitle=\bfseries,
    boxrule=0.8pt,
    arc=2pt,
    left=6pt, right=6pt, top=6pt, bottom=6pt,
    breakable
}{thm}

\newtcolorbox{myrem}[1][]{
    colback=remarkgray,
    colframe=remarkborder,
    fonttitle=\bfseries,
    title=#1,
    boxrule=0.6pt,
    arc=2pt,
    left=6pt, right=6pt, top=4pt, bottom=4pt,
    breakable
}

\AtBeginDocument{%
  }

\setcopyright{none}
\acmDOI{}
\acmPrice{}

\makeatletter
\AtBeginDocument{%
  \ifx\@acmDOI\@empty\let\@formatdoi\@gobble\fi}
\makeatother

\begin{document}
\title[MADRec: Reliable Multi-Agent Reasoning for Recommendation]{MADRec: Reliable Multi-Agent Reasoning with Evidence-Grounded Debate for Recommendation}

\author{Zunlong Liu}
\email{cpuzzzq@163.com}
\orcid{0009-0009-1685-8099}
\affiliation{
  \institution{Hefei University of Technology}
 \city{Hefei}
  \country{China}
}

\author{Weijie Feng}
\authornote{Corresponding Authors.}
\orcid{0009-0005-1911-4790}
\email{wjfeng@hfut.edu.cn}
\affiliation{
  \institution{Hefei University of Technology}
 \city{Hefei}
  \country{China}
}

\author{Jingzhi Li}
\email{jingzhili826@gmail.com}
\orcid{0009-0005-5094-1786}
\affiliation{
  \institution{Hefei University of Technology}
  \city{Hefei}
  \country{China}
}

\author{Lei Zhu}
\email{leizhu0608@gmail.com}
\orcid{0000-0002-2993-7142}
\affiliation{
  \institution{Tongji University}
 \city{Shanghai}
  \country{China}
}

\author{Zhiyong Cheng}
\authornotemark[1]
\orcid{0000-0003-1109-5028}
\email{jason.zy.cheng@gmail.com}
\affiliation{
  \institution{Hefei University of Technology}
 \city{Hefei}
  \country{China}
}

\author{Meng Wang}
\email{eric.mengwang@gmail.com}
\orcid{0000-0002-3094-7735}
\affiliation{
  \institution{Hefei University of Technology}
 \city{Hefei}
  \country{China}
}

\renewcommand{\shortauthors}{Liu et al.}


\begin{abstract}

Large Language Models (LLMs) are transforming recommender systems from conventional preference matching toward reasoning-driven and explainable decision-making. Multi-agent collaboration further advances this paradigm by distributing recommendation reasoning across specialized agents. However, existing frameworks rarely assess the reliability of intermediate reasoning during collaboration, allowing erroneous judgments to propagate across agents and ultimately degrade recommendation quality. To address this issue, we propose \textbf{\underline{M}ulti-\underline{A}gent \underline{D}ebate-based \underline{Rec}ommendation (MADRec)}, an evidence-grounded multi-agent framework that enhances the reliability of multi-agent recommendation reasoning. 
MADRec consists of three coordinated modules. First, the \textbf{Multi-Agent Debate (MAD)} module enables an Interaction Agent and a Semantic Agent to analyze interaction signals and content-level semantic signals from complementary perspectives, generating diverse recommendation judgments through structured debate. Second, the \textbf{Evidence-Grounded Retrieval-Augmented
Reasoning (EG-RAR)} module grounds the Semantic Agent's
reasoning in item-level evidence and refines candidate
scores by assessing how well the evidence supports
the corresponding semantic claims. Third, the \textbf{Judge-based Ranking Calibration (JRC)} module leverages debate-derived reliability cues to adaptively calibrate interaction-based and semantic ranking results for the final recommendation. Extensive experiments on three real-world datasets demonstrate that MADRec consistently outperforms competitive baselines and enhances the reliability of multi-agent recommendation reasoning.
\end{abstract}



\maketitle

\input{sect1_intro}

\input{sect2_relatedwork_revised}

\input{sect3_method_revised}
\input{sect4_experiment}
\input{sect5_conclution}
\input{sect6_acknowledgments}

\bibliographystyle{ACM-Reference-Format}
\bibliography{references}

\clearpage
\appendix
\input{sect7_appendix}

\end{document}

%% file: sect1_intro.tex
\section{Introduction}
Recent advances in Large Language Models (LLMs)~\cite{achiam2023gpt,liu2024deepseek,liu2025arts} have stimulated interest in agent-based recommender systems~\cite{toolrec,recagent}. Unlike traditional models that generate rankings by learning latent preference representations from user-item interactions~\cite{s3rec,sun2019bert4rec,liu2026immpoi,PP}, LLM-based agents introduce explicit reasoning into recommendation and leverage capabilities such as planning, memory, and tool use~\cite{introplan1,intromemory1,introtool1} to reason over interaction patterns, item semantics, and contextual information. This shift moves recommender systems from conventional preference matching toward reasoning-driven and explainable decision-making, where recommendation quality increasingly depends on the reliability of agent reasoning.

Existing research on agent-based recommendation can be broadly categorized into two paradigms~\cite{peng2025survey,loopagentic}. Simulation-oriented methods~\cite{agent4rec,recllm,recagent,zhang2024agentcf} employ user agents or interaction simulators to model behaviors and interaction dynamics, whereas recommendation-oriented methods leverage LLM-based agents to improve recommendation decisions through preference reasoning and candidate evaluation. Early recommendation-oriented methods~\cite{interecagent,wang2024recmind,wang2023drdt} typically rely on a single agent to perform the entire reasoning and decision-making process. However, recommendation decisions often depend on diverse information signals, including historical interaction patterns, item semantics, and contexts. Giving all these signals to a single agent can lead it to focus
too heavily on some preference cues and overlook others, making
it harder to fully model user preferences~\cite{chen2024reconcile}. To address this limitation, recent studies have explored multi-agent recommendation frameworks~\cite{wang2024macrec,shu2024rah}, which distribute recommendation reasoning across specialized agents to better capture heterogeneous preference signals.

\begin{figure}[t]
  \centering
  \includegraphics[width=0.85\linewidth]{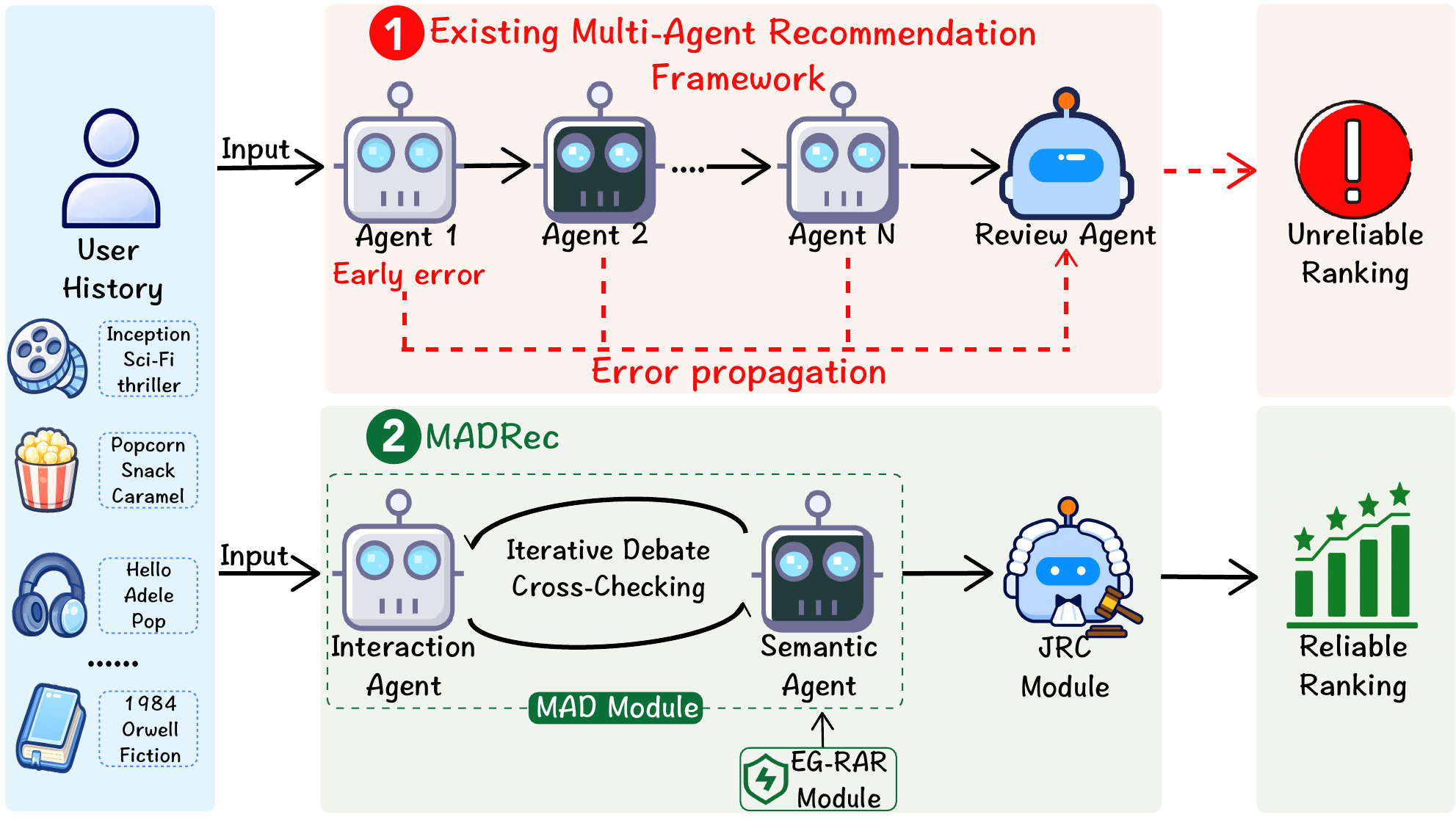}
  \Description{Comparison of how an intermediate ranking error propagates through conventional agent collaboration and how MADRec reassesses the error before final ranking.}
  \vspace{-10pt}
  \caption{Illustration of error propagation and its mitigation.}
  \vspace{-10pt}
  \label{fig:intro}
\end{figure}

In existing multi-agent recommendation frameworks, different stages or subtasks of the recommendation process are typically assigned to multiple role-specific agents~\cite{wang2024macrec,shu2024rah,macrs,banerjee2025collab}. Each agent performs reasoning over different aspects of the recommendation task, and its intermediate reasoning is subsequently incorporated into downstream agents' reasoning contexts~\cite{introsurvey1,introsuvery2}. The downstream agent may rely on the previous agent's judgment
without checking the candidate again. If the earlier judgment is
wrong, the downstream agent may make further errors that can
affect the final recommendation ranking~\cite{Survey1,introsuvery2}.
As shown in Figure~\ref{fig:intro}, \textit{error propagation}
occurs when an agent's incorrect judgment influences other
agents and affects the final recommendation ranking
~\cite{errorpropagation}. Although some methods introduce review, reflection, or coordination components~\cite{macrs,wang2024macrec,macf,sun2026pfmar}, these mechanisms are typically applied after most reasoning steps have been completed. Consequently, they may fail to prevent unreliable intermediate reasoning from influencing downstream agents during the collaborative reasoning process. This motivates mechanisms that allow agents to review earlier judgments
about candidates during collaboration and reduce the impact of errors on
later rankings.


Multi-agent debate offers a promising process-level mechanism for
mitigating this issue. Through multi-round interactions, agents can
compare and challenge each other's intermediate judgments, exposing
potential inconsistencies and enabling iterative
refinement~\cite{du,chan2024chateval,jeong2026toolmad}. This mechanism
is particularly suitable for recommendation, where final decisions
often require integrating multiple preference signals that reflect
different aspects of user interests. Such diversity helps agents form complementary perspectives, enabling agents to
cross-check intermediate judgments and expose potentially unreliable
decisions. 

Recent studies have explored multi-agent debate in recommender systems~\cite{eyecrs,li2026debate4rec,crave}. However, using debate to reduce error propagation during collaborative
ranking requires agents to reassess candidates based on each other's
feedback and then review
the revised rankings in later rounds to correct the positions of
individual items~\cite{du2024improving,xie2026spark,lin2026agentask}. In CRAVE's evaluated setting~\cite{crave}, one-way critique
allows the later agent to revise the recommendation after
reviewing an earlier judgment, leaving the earlier agent no
further turn to reconsider its judgment or check the revision. Debate4Rec~\cite{li2026debate4rec} uses debate to select a final recommendation list from several generated lists. However, selection alone cannot correct item-ranking errors shared by all available lists. Even if agents identify errors during debate, they may still rely on unsupported claims about item content or attributes when revising their judgments~\cite{makde,zhou2026guardian}. If other agents accept these claims, the resulting errors may affect later candidate assessments and ranking decisions. Reducing error propagation therefore requires both reciprocal checking and revision across agents and an assessment of whether item-level evidence supports the underlying semantic claims.

Inspired by the above insights, we propose \textbf{MADRec}, an evidence-grounded multi-agent recommendation framework that aims to mitigate error propagation in collaborative recommendation reasoning. MADRec follows a two-stage retrieval-and-ranking pipeline, where the retrieval stage first retrieves a user-specific candidate set, and the ranking stage improves the reliability of agent reasoning through three coordinated modules: \textbf{Multi-Agent Debate (MAD)}, \textbf{Evidence-Grounded Retrieval-Augmented Reasoning (EG-RAR)}, and \textbf{Judge-based Ranking Calibration (JRC)}.
Specifically, \textbf{MAD} introduces an \textit{Interaction Agent} and a \textit{Semantic Agent} that assess the same candidate set using interaction signals and item semantics, respectively. Through reciprocal feedback across debate rounds, each agent can challenge the other’s judgments about individual candidates and revise its own assessments and ranking in response to the critiques it receives. However, debate alone cannot guarantee that semantic judgments are sufficiently supported by item-level evidence. Therefore, \textbf{EG-RAR} grounds the \textit{Semantic Agent}'s reasoning in retrieved item-level evidence and refines candidate scores
based on how well the evidence supports the semantic claims about each candidate. The two agents may still rank candidates differently after debate because they use different preference signals. Therefore, \textbf{JRC} leverages reliability cues derived from agent interactions to calibrate  interaction-based and semantic ranking outputs, producing a unified recommendation ranking. By integrating collaborative debate, evidence grounding, and reliability-aware calibration, MADRec reduces the impact of unreliable intermediate reasoning and improves the robustness of multi-agent recommendation.
We have released the code and relevant parameter settings to facilitate
repeatability as well as further research. 


In summary, our contributions are threefold:
\begin{itemize}[leftmargin=0.35cm,topsep=3pt]
\item  We propose MADRec, an evidence-grounded multi-agent recommendation framework that mitigates
error propagation in collaborative agent reasoning. Built upon a two-stage retrieval-and-ranking pipeline, MADRec enhances the reliability of multi-agent reasoning in recommender systems.

\item We develop three complementary modules, namely \textbf{MAD}, \textbf{EG-RAR}, and \textbf{JRC}. \textbf{MAD} enables agents to exchange feedback and refine their ranking decisions through multi-round debate. \textbf{EG-RAR} uses item-level evidence to evaluate semantic claims and adjust their scores based on evidence support. \textbf{JRC} assigns adaptive weights to the interaction and semantic branches to produce a more reliable final ranking.

\item Extensive experiments on three real-world datasets demonstrate that MADRec outperforms competitive baselines. Ablation studies further support the effectiveness of the individual modules. Controlled perturbations of intermediate rankings show that MADRec reduces error transfer between agents, improves the correction rate of transferred errors, and reduces error persistence in the final ranking.
\end{itemize}

%% file: sect2_relatedwork_revised.tex
\section{Related Work}
\label{Related_work}
\subsection{Agent-based Recommendation}
Agent-based recommendation frameworks leverage the planning, reasoning, and task-execution capabilities of LLMs to support recommendation related tasks~\cite{wang2023drdt,billp,wang2025re2llm}. Existing studies can be broadly categorized into user simulation-oriented and recommendation task-oriented methods. The former employ LLM agents to simulate user behaviors, interaction dynamics, and preference evolution through profile, memory, and action modeling~\cite{agent4rec,recllm,recagent,zhang2024agentcf}, primarily supporting simulation and analysis rather than directly improving recommendation decisions.

Recommendation-oriented methods directly employ LLM-based agents for recommendation. Early studies typically use a single agent for reasoning, planning, tool use, or reflection~\cite{interecagent,wang2024recmind}. For example, InteRecAgent~\cite{interecagent} integrates memory
and recommendation tools to support task planning and execution.
RecMind~\cite{wang2024recmind} explores agentic reasoning and
tool use for recommendation, while ToolRec~\cite{toolrec}
uses tools to retrieve and evaluate items. These methods enable
an agent to gather information and refine recommendation
decisions beyond a single prediction step.

Multi-agent approaches further distribute recommendation
reasoning across specialized roles or perspectives.
MACRec~\cite{wang2024macrec} employs a Manager to coordinate
User/Item Analysts, a Reflector, and other agents for different
recommendation tasks. The Reflector provides feedback that
allows the Manager to revise previous outputs.
MACF~\cite{macf} instantiates similar users and relevant items
as agents and uses an orchestrator to organize multi-round
collaboration, incorporating neighborhood-based collaborative
signals into recommendation reasoning.

These methods demonstrate the value of specialized analysis,
iterative feedback, and agent coordination. Building on this
direction, MADRec focuses on the reliability of intermediate
candidate judgments exchanged between interaction-based and
semantic perspectives. It combines reciprocal feedback over
a shared candidate set with explicit claim--evidence support
assessment and adaptive aggregation of the resulting rankings.

\subsection{Multi-Agent Debate for Recommendation}
Multi-agent debate enables LLM agents to examine and refine their judgments through argument exchange and mutual critique. Existing frameworks explore diverse perspectives, iterative feedback, and aggregation mechanisms to improve reasoning quality~\cite{du2024improving,liang2024encouraging,shinn2023reflexion,chen2024reconcile,jeong2026toolmad,makde}. Recent studies have further extended this paradigm to recommendation tasks, where decisions depend on personalized user preferences and contextual information.

CRAVE~\cite{crave} proposes an experience-augmented conversational recommendation framework that leverages verbalized experiences from historical dialogues and user feedback to guide a Debater-Critic process. However, its evaluated one-round configuration follows a unidirectional feedback process, preventing earlier debaters from revisiting their judgments in response to subsequent critiques. This asymmetric interaction may limit mutual error correction and make the resolution of erroneous judgments more dependent on downstream agents. EyeCRS~\cite{eyecrs} employs supporting and opposing agents to identify historical dialogue information that may cause negative transfer. However, its debate focuses on evaluating historical dialogue information rather than directly refining candidate-level recommendation judgments. Debate4Rec~\cite{li2026debate4rec} addresses the trade-off between preference alignment and content diversity in generative recommendation, using debate-based feedback to select among generated recommendation lists. However, the selected
list retains its original item order. If every candidate list
ranks a relevant item too low, even the best list will retain
that error. Correcting such an error requires agents to revisit
their assessment of the item and revise its position.

In contrast, MADRec focuses on the reliability of intermediate
item-level judgments during recommendation. It combines
reciprocal feedback between interaction-based and semantic
perspectives with evidence support assessment, allowing
evidence-adjusted judgments to inform subsequent agent
interaction and final ranking.

\begin{figure}[htbp]
\centering
\includegraphics[width=1.0\textwidth]{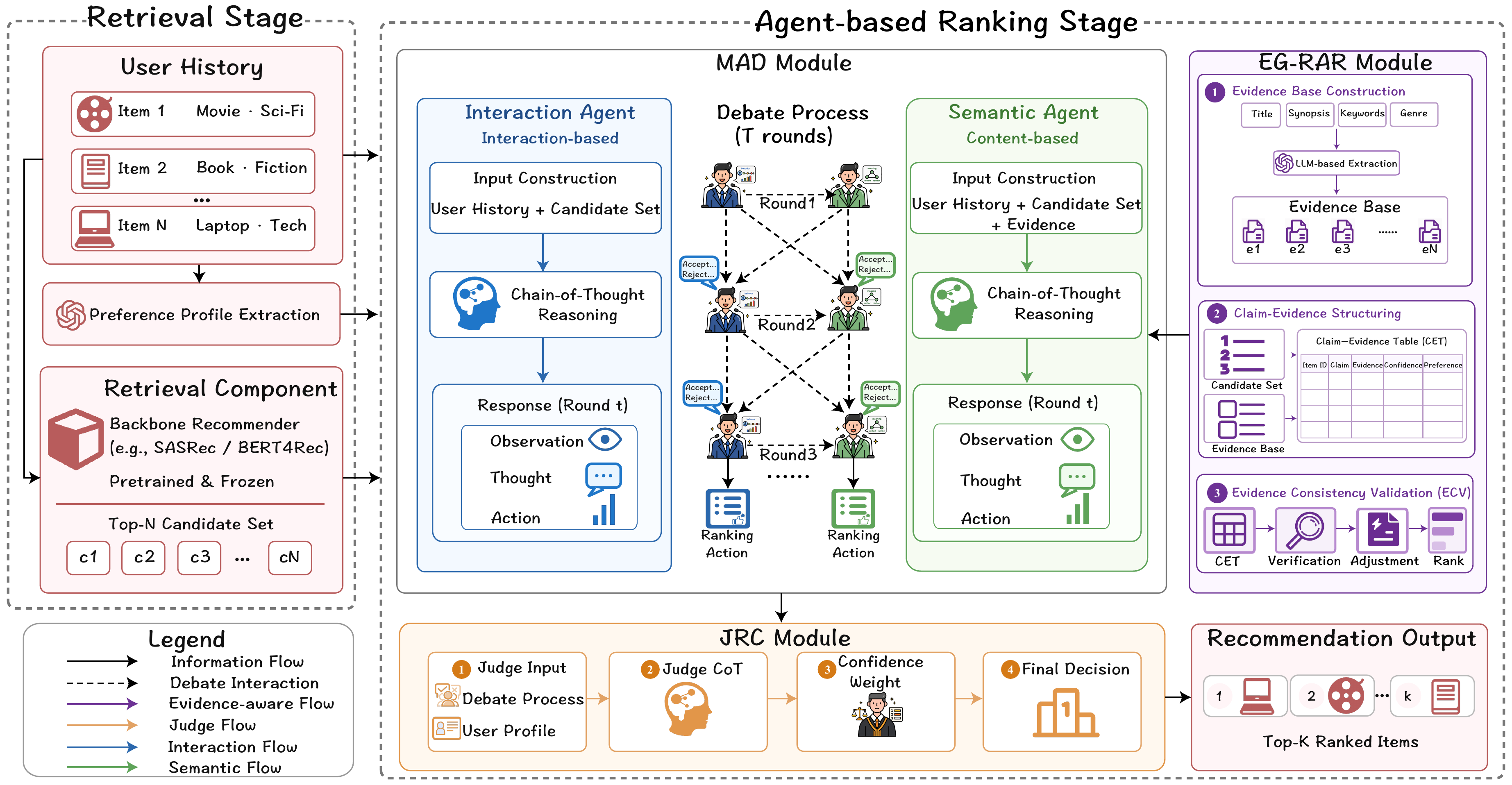}
\Description{The retrieval stage produces candidates. Interaction and Semantic Agents debate the candidate rankings, the Semantic Agent verifies item evidence, and the Judge Agent calibrates the final ranking.}
\vspace{-20pt}
\caption{Overall framework of MADRec. It consists of two stages: retrieval and agent-based ranking. MAD enables multi-agent debate, EG-RAR provides item-level evidence, and JRC calibrates the final ranking.}
\vspace{-20pt}
\label{fig:framework}
\end{figure}


%% file: sect3_method_revised.tex
\section{Methodology}
\subsection{Overview}
\label{sec:overview}
As illustrated in Figure~\ref{fig:framework}, MADRec follows a two-stage retrieval-and-ranking pipeline consisting of candidate retrieval and agent-based ranking. In the retrieval stage, MADRec first retrieves a user-specific candidate set from the full item space. The ranking stage then refines these candidates through three coordinated modules: \textit{Multi-Agent Debate} (MAD), \textit{Evidence-Grounded Retrieval-Augmented Reasoning} (EG-RAR), and \textit{Judge-based Ranking Calibration} (JRC). These modules respectively enhance recommendation reliability at three levels: interaction-level cross-checking, evidence-level grounding, and decision-level calibration.

\textbf{Problem Formulation.}
MADRec is formulated under the general recommendation setting. Given a user $u$ with historical interaction records $\mathcal{H}_{u}=\{i_1,i_2,\ldots,i_{n-1}\}$, where $i_j$ denotes an item previously interacted with by the user, the goal is to recommend a ranked list of items that best match the user's preferences from the complete item set $\mathcal{I}$.

\subsection{Two-Stage Recommendation Pipeline}
\label{sec:pipeline}


To support scalable recommendation over the large item space, MADRec adopts a two-stage retrieval-and-ranking pipeline consisting of a \textbf{retrieval stage} and an \textbf{agent-based ranking stage}. The \textit{retrieval stage} first retrieves a user-specific candidate set from the full item space, while the \textit{agent-based ranking stage} performs agent-based reasoning over the retrieved candidates and produces the final recommendation ranking.

\subsubsection{Retrieval Stage} 
For each user $u$, this retrieval stage retrieves a user-specific candidate set $C_u$ from the full item space based on a pretrained recommendation model. Specifically, given a retrieval model $f_{ret}(\cdot)$, the retrieval component estimates the relevance between the user and candidate items and selects the top-$N$ items as the candidate set:
\begin{equation} 
C_u=TopN_{i\in \mathcal{I}}s_{ret}(u,i),
\end{equation} 
where $s_{ret}(u,i)$ denotes the relevance score produced by the retrieval model. The retrieved candidates are then passed to the Ranking Component for agent-based reasoning and final ranking.

\subsubsection{Agent-based Ranking Stage}
\label{sec:ranking}
As described above, the agent-based ranking stage consists of three coordinated modules: \textbf{MAD}, \textbf{EG-RAR}, and \textbf{JRC}. Conditioned on historical interaction records $\mathcal{H}_u$, user profile $P_u$\footnote{The user profile refers to an implicit representation of user preferences inferred from historical interactions using LLMs.}, and candidate set $C_u$, these modules perform reasoning and ranking over the retrieved candidates.

First, \textbf{MAD} introduces two specialized agents, namely an \textit{Interaction Agent} $\pi_I$ and a \textit{Semantic Agent} $\pi_S$, to reason about candidate items from complementary perspectives. $\pi_I$ derives preference judgments from user interaction signals, whereas $\pi_S$ analyzes item semantics and latent user intents from a content-oriented perspective. Through multi-round debate, the two agents exchange and challenge their intermediate judgments, allowing them to examine different reasoning perspectives and identify potentially inconsistent judgments during collaborative recommendation.

Second, \textbf{EG-RAR} is introduced to further improve the reliability of semantic reasoning. Although debate
enables cross-agent examination, semantic judgments may still lack sufficient support from item-level evidence.
\textbf{EG-RAR} therefore retrieves relevant evidence for the Semantic Agent and organizes its semantic claims and the corresponding evidence into a claim-evidence structure. It then estimates the extent to which the evidence supports each claim and uses this assessment to adjust the corresponding candidate score. These adjusted scores then affect semantic ranking and subsequent agent interactions.

Finally, \textbf{JRC} performs decision-level calibration over the ranking outputs from different reasoning perspectives. Since interaction-based and semantic reasoning may have different reliability under different recommendation contexts, JRC leverages accumulated debate information to estimate their relative reliability and adaptively calibrate their contributions to the final recommendation ranking.

Detailed descriptions of these components are provided in the following subsections, and the prompt templates used in these modules are provided in Appendix~\ref{app:details}.

\subsection{Multi-Agent Debate Module}
The \textbf{MAD} module is designed to provide interaction-level
cross-checking during multi-agent collaboration. Its key idea is to enable the \textit{Interaction Agent} $\pi_I$ and the \textit{Semantic Agent} $\pi_S$ to examine each other's intermediate reasoning and ranking outputs, allowing inconsistent or potentially
unreliable judgments to be exposed during agent interaction. The two agents follow a structured reasoning-trace generation paradigm inspired by Chain-of-Thought (CoT) prompting~\cite{cot}, where intermediate reasoning traces are maintained to support subsequent debate actions. The debate process follows an \textbf{\textit{observation} $\rightarrow$ \textit{think} $\rightarrow$ \textit{action}} loop,
where each agent first observes the counterpart's intermediate outputs, then generates a reasoning trace $g$ to analyze its own judgments and critique the counterpart's outputs, and finally produces a ranking action $A$ based on the updated reasoning context.

\paragraph{\textbf{Observation} (Cross-Agent Feedback).}
In MADRec, agents do not receive direct external feedback as in interactive agent environments~\cite{yao2022react}. Instead, each agent treats the other agent's reasoning trace and ranking action as cross-agent observations for subsequent reasoning. Since the \textit{Interaction Agent} $\pi_{I}$ acts before the \textit{Semantic Agent} $\pi_{S}$ in each debate round, its observation at round $t$ is defined as $o_I^{(t)}=(g_S^{(t-1)},A_S^{(t-1)})$, whereas the \textit{Semantic Agent} observes the \textit{Interaction Agent}'s current output, i.e., $o_S^{(t)}=(g_I^{(t)},A_I^{(t)})$. For the first round, the \textit{Interaction Agent} starts without cross-agent feedback, i.e., $o_I^{(1)}=\emptyset$. These observations are incorporated into the corresponding reasoning contexts, enabling each agent to consider the opponent's judgments during subsequent reasoning.

\paragraph{\textbf{Think} (Reasoning).}
The \textit{think} step generates intermediate reasoning traces that summarize each agent's preference analysis and critique of the opponent's judgments. At each debate round, the reasoning process is conditioned on two types of information: cross-agent observations received from the opponent and the agent's own reasoning traces and ranking actions. The cross-agent observations provide external perspectives on candidate evaluation, while the accumulated reasoning history enables each agent to track previous decisions and adjust subsequent judgments. Accordingly, the interaction contexts of the two agents are defined as:
\begin{equation}
\begin{aligned}
c_I^{(t)}
&=
\left(
\{o_S^{(\tau)}\}_{\tau=1}^{t-1},
\{o_I^{(\tau)}\}_{\tau=1}^{t}
\right), 
c_S^{(t)}=
\left(
\{o_I^{(\tau)}\}_{\tau=1}^{t},
\{o_S^{(\tau)}\}_{\tau=1}^{t}
\right).
\end{aligned}
\end{equation}

Based on the interaction context, the \textit{Interaction Agent} and \textit{Semantic Agent} generate reasoning traces at round $t$:
\begin{equation}
\begin{aligned}
g_{I}^{(t)}
=
\mathrm{CoT}_{I}\left(
\mathcal{H}_{u},P_u,
C_{u},
c_{I}^{(t)}
\right),
g_{S}^{(t)}
=
\mathrm{CoT}_{S}\left(
\mathcal{H}_{u},P_{u},
C_{u},
c_{S}^{(t)}
\right).
\end{aligned}
\label{gongshi:thought}
\end{equation}

Here, $\mathrm{CoT}_{I}(\cdot)$ and $\mathrm{CoT}_{S}(\cdot)$ denote the role-specific prompting procedures for generating reasoning traces of the \textit{Interaction Agent} and the \textit{Semantic Agent}, respectively. The prompt templates for these reasoning instructions are provided in Appendix~\ref{app:details}.

The generated reasoning traces contain both preference analysis and critique information about the opponent agent's judgments. Such cross-agent feedback is maintained within the reasoning context and guides subsequent debate rounds.


\paragraph{\textbf{Action} (Ranking).}
The \textit{action} step converts each agent's reasoning trace into a ranking action over the candidate set. Given the reasoning trace $g_I^{(t)}$, the \textit{Interaction Agent} $\pi_I$ produces item-level preference scores and ranking rationales: $(p_{I,i}^{(t)},q_{I,i}^{(t)})=\pi_I(g_I^{(t)},i)$ for $i\in C_u$, where $p_{I,i}^{(t)}\in[0,1]$ denotes the preference score and $q_{I,i}^{(t)}$ provides the corresponding reasoning explanation based on interaction signals. It then generates the ranking action:
\begin{equation}
    A_I^{(t)}=[(i,p_{I,i}^{(t)},q_{I,i}^{(t)})]_{i\in\operatorname{Top-K}_{j\in C_u}(p_{I,j}^{(t)})}.
\end{equation}

For the \textit{Semantic Agent}, the ranking action $A_S^{(t)}$ is generated based on its semantic reasoning trace $g_S^{(t)}$, with evidence grounding incorporated through \textbf{EG-RAR} as described in Section~\ref{sec:EG-RAR}.

\paragraph{\textbf{Full Debate Process}.}

The debate begins in \textit{Round 1}, where the \textit{Interaction Agent} $\pi_I$ generates an initial reasoning trace $g_I^{(1)}$ and ranking action $A_I^{(1)}$ from the user’s interactions $\mathcal{H}{u}$, profile $P{u}$, and candidate set $C_{u}$. The \textit{Semantic Agent} $\pi_S$ then observes these outputs and generates reasoning trace $g_S^{(1)}$ and ranking action $A_S^{(1)}$.

For subsequent rounds $(t \geq 2)$, both agents iteratively update their reasoning traces and ranking actions using accumulated interaction contexts derived from cross-agent observations. During each round, agents compare the opponent's reasoning and ranking outputs with their own judgments and refine their subsequent decisions based on the observed feedback. After a fixed number of rounds $T$, the resulting interaction-based and semantic ranking outputs are passed to \textbf{JRC} for final calibration.

\subsection{Evidence-Grounded Retrieval-Augmented Reasoning Module}
\label{sec:EG-RAR}

The \textbf{MAD} module enables agents to cross-check intermediate judgments from complementary perspectives. However, debate mainly improves reasoning consistency and cannot guarantee that semantic judgments are grounded in reliable item-level evidence. The \textit{Semantic Agent }may still generate weakly supported claims from incomplete or ambiguous item information. To address this limitation, \textbf{EG-RAR} grounds semantic
reasoning in item-level evidence and incorporates evidence
support into candidate scoring. Specifically, it constructs
an evidence base for retrieving relevant item information
and organizes semantic claims and their associated evidence
into a \textit{Claim-Evidence Table} (CET). An \textit{Evidence Consistency Validator} (ECV) then
estimates how well the evidence supports each claim
and uses the resulting support coefficient to adjust
the corresponding semantic ranking score.

\subsubsection{Evidence Base Construction}
\label{sec:evidence_base}
To support evidence-grounded semantic reasoning, we construct an item-level evidence base offline. For each item $i \in \mathcal{I}$, we collect item-related information from available structured and unstructured sources, where the collected information contains semantic descriptions and item attributes. The collected item information is denoted as $x_i$.
We then employ an LLM with a predefined semantic schema prompt $\mathcal{S}$ to transform the raw item information into structured semantic evidence: $\phi_i=\mathrm{LLM}(x_i;\mathcal{S})=(S_i,F_i,C_i)$, where $S_i$, $F_i$, and $C_i$ denote semantic, functional, and contextual characteristics, respectively. 

For each item, the evidence entry is defined as $\epsilon_i=(i,x_i,\phi_i)$. All item-level evidence entries form the evidence corpus: $\mathcal{E}={\{\epsilon_i\mid i\in\mathcal{I}}\}$, which is retrieved during online semantic reasoning to provide explicit item-level support for candidate evaluation. The structured representation $\phi_i$ provides concise semantic
information for recommendation reasoning, while the original
item information $x_i$ is retained as the source of evidence
for claim verification.
\subsubsection{Claim-Evidence Table}
\label{sec:cet}

During online inference, given the candidate set $C_u$, \textbf{EG-RAR}
retrieves item-level evidence entries from the evidence corpus $\mathcal{E}$
for each candidate item. The retrieved evidence is incorporated into the
\textit{Semantic Agent}'s reasoning context, transforming the original semantic reasoning process into evidence-grounded reasoning. 
Based on the retrieved evidence, the \textit{Semantic Agent} generates an evidence-aware reasoning trace $g_S^{(t)}$ and automatically constructs a structured CET. At debate round $t$, the CET is defined as:
\begin{equation}
\begin{aligned}
\mathrm{CET}^{(t)}
&=
\left\{
m_i^{(t)}
\mid i\in C_u
\right\},m_i^{(t)}
=
\left(
i,
q_{S,i}^{(t)},
d_i^{(t)},
p_{S,i}^{(t)},
\rho_{S,i}^{(t)}
\right),
\end{aligned}
\label{eq:cet}
\end{equation}
Here, $q_{S,i}^{(t)}$ denotes an item-related factual claim
used to support the Semantic Agent's preference assessment,
and $d_i^{(t)}$ denotes the corresponding evidence passage
from the original item information $x_i$.
The preference score $p_{S,i}^{(t)}$ and confidence score
$\rho_{S,i}^{(t)}$ represent the Semantic Agent's assessment
of the item's relevance to the user and its confidence in
that assessment, respectively.

\subsubsection{Evidence-Aware Score Calibration}
\label{sec:ecv}

Building on the claim–evidence pairs organized in the CET, we introduce the ECV to assess how well the associated evidence supports the factual content of each semantic claim. The resulting support estimates are used to adjust semantic ranking scores, allowing the contribution of semantic judgments to reflect their evidential support.

For each record $m_i^{(t)} \in \mathrm{CET}^{(t)}$,
ECV takes the semantic claim $q_{S,i}^{(t)}$
and its corresponding evidence $d_i^{(t)}$ as input. We formulate verification as a binary classification task
with label set
$\mathcal{Y}=\{\mathrm{Supported},\mathrm{NotSupported}\}$.
The label $\mathrm{Supported}$ indicates that all factual claims are supported by the evidence, while $\mathrm{NotSupported}$ indicates that at least one claim is unsupported or contradicted. ECV outputs a probability distribution over the two labels:
\begin{equation}
\label{eq:verifier}
\mathbf{P}_i^{(t)}
=
\begin{bmatrix}
V_{\psi}(\mathrm{Supported}\mid d_i^{(t)},q_{S,i}^{(t)})\\
V_{\psi}(\mathrm{NotSupported}\mid d_i^{(t)},q_{S,i}^{(t)})
\end{bmatrix}.
\end{equation}
Here, $V_{\psi}$ denotes the ECV model with parameters $\psi$. Given evidence $d_i^{(t)}$ and claim $q_{S,i}^{(t)}$, it predicts the probability of each label $y\in\mathcal{Y}$.

We implement $V_{\psi}$ using a pretrained DeBERTaV3-base encoder\footnote{https://huggingface.co/microsoft/deberta-v3-base} with a binary classification head. The model is fine-tuned on $\mathcal{D}_{\mathrm{ver}}$, the
pooled ECV training set of labeled evidence-claim pairs.
Appendix~\ref{app:ecv} describes how these pairs are
constructed and labeled. Each training example is represented as $\left(d_i^{(t)}, q_{S,i}^{(t)}, y_i^{(t)}\right)$, where $y_i^{(t)} \in \mathcal{Y}$ denotes the support label. The encoder and classification head are jointly optimized using the cross-entropy loss:
\begin{equation}
\label{eq:verifier_loss}
\mathcal{L}_{\mathrm{ver}}
=
-\frac{1}{|\mathcal{D}_{\mathrm{ver}}|}
\sum_{\left(d_i^{(t)},q_{S,i}^{(t)},y_i^{(t)}\right)
\in \mathcal{D}_{\mathrm{ver}}}
\log V_{\psi}\left(
y_i^{(t)} \mid d_i^{(t)},q_{S,i}^{(t)}
\right).
\end{equation}

After training, ECV is frozen during recommendation inference. Its predicted probability of \(\mathrm{Supported}\) is used as the evidence support coefficient:
\begin{equation}
\label{eq:support_coefficient}
\gamma_i^{(t)}
= V_{\psi}\left(
\mathrm{Supported}
\mid d_i^{(t)}, q_{S,i}^{(t)}
\right).
\end{equation}

The adjusted semantic score is computed as:
\begin{equation}
\label{eq:adjusted_semantic_score}
\widetilde{\eta}_i^{(t)}
= p_{S,i}^{(t)}
\rho_{S,i}^{(t)}
\gamma_i^{(t)},
\end{equation}
where $p_{S,i}^{(t)}$ and $\rho_{S,i}^{(t)}$ denote the Semantic Agent's preference score and self-reported confidence, respectively. The coefficient $\gamma_i^{(t)}$ adjusts the semantic score based on the evidence support for the claim. Candidates are ranked in descending order of $\widetilde{\eta}_i^{(t)}$, yielding the semantic Top-$K$ ranking:
\begin{equation}
\label{eq:semantic_ranking}
R_S^{(t)}
=
\left[
\left(
i,\widetilde{\eta}_i^{(t)},q_{S,i}^{(t)}
\right)
\right]_{
i\in\operatorname{TopK}_{j\in C_u}
\left(\widetilde{\eta}_j^{(t)}\right)
}.
\end{equation}
The adjusted semantic ranking is returned as $A_S^{(t)}=R_S^{(t)}$ for subsequent agent interactions. The final-round adjusted scores are also used by JRC for ranking aggregation.

\subsection{Judge-based Ranking Calibration Module}
\label{sec:judge}
After $T$ rounds of debate, the \textit{Interaction Agent} and the
\textit{Semantic Agent} produce final ranking outputs from two
complementary perspectives: \textit{interaction-driven preference}
and \textit{item-level semantic relevance}. Although these outputs
have been examined through multi-agent debate, they remain
perspective-specific and may exhibit different levels of reliability
under different recommendation contexts. To perform decision-level
calibration, MADRec introduces the \textbf{JRC} module to analyze
the debate outcomes, estimate the relative reliability of the two
perspectives, and generate the final recommendation ranking.

Specifically, \textbf{JRC} employs a \textit{Judge Agent} $\pi_J$
as a decision-level calibrator. Unlike the specialized \textit{Interaction}
and \textit{Semantic} agents, the \textit{Judge Agent} jointly considers
user history $\mathcal{H}_u$, user profile $P_u$, and accumulated
debate history. The debate history after $T$ rounds is denoted as
$\mathcal{D}^{(T)}
=
\{(g_I^{(t)},A_I^{(t)},g_S^{(t)},A_S^{(t)})\}_{t=1}^{T}$.

\textbf{JRC} first constructs a calibration candidate set from the
union of the two agents' final Top-$K$ ranking actions:
\begin{equation}
C_u^{J}
=
\operatorname{Items}\!\left(A_I^{(T)}\right)
\cup
\operatorname{Items}\!\left(A_S^{(T)}\right),
\end{equation}
where $\operatorname{Items}(A)$ extracts candidate items from a ranking action $A$. The union of the two final Top-$K$ ranking actions forms the calibration candidate set. Since both agents evaluate all retrieved candidates before selecting Top-$K$ items, their corresponding final-round scores remain available for all items in $C_u^J$.


Based on the user information, accumulated debate history, and
calibration candidate set, the \textit{Judge Agent} generates a
calibration reasoning trace
$g_J^{(T)}
=
\mathrm{CoT}_J(\mathcal{H}_u,P_u,\mathcal{D}^{(T)},C_u^{J})$,
where $\mathrm{CoT}_{J}(\cdot)$ denotes the \textit{Judge Agent}'s
prompting procedure. The complete prompt template is provided in
Appendix~\ref{app:details}. Based on the calibration reasoning trace, the \textit{Judge Agent} infers reliability weights $\omega_I^{(T)}$ and $\omega_S^{(T)}$ for the interaction-based and semantic perspectives, respectively. These weights reflect the consistency and supporting signals of different reasoning perspectives captured in the accumulated debate history, satisfying
$\omega_I^{(T)}+\omega_S^{(T)}=1$ and
$\omega_I^{(T)},\omega_S^{(T)}\in[0,1]$.

For each such item, \textbf{JRC} combines the final-round interaction-based preference score and the evidence-aware semantic score produced by \textbf{EG-RAR}:
\begin{equation}
s_{J,i}
=
\omega_I^{(T)}p_{I,i}^{(T)}
+
\omega_S^{(T)}\widetilde{\eta}_i^{(T)},
\quad i\in C_u^{J}.
\end{equation}
Here, $s_{J,i}$ denotes the final calibrated score assigned to item
$i$. The final list is obtained as
$A^*=\operatorname{Top-K}_{i\in C_u^{J}}(s_{J,i})$.

%% file: sect4_experiment.tex
\section{Experiments}
\subsection{Experimental Setup}
\label{Setup}
MADRec is formulated as a general recommendation ranking framework that operates on retrieved candidate sets without relying on a specific retrieval architecture. To evaluate its effectiveness in a practical next-item recommendation scenario, we instantiate MADRec in the sequential recommendation setting in this work, where pretrained sequential recommenders are adopted as retrieval backbones for candidate generation. The retrieval component generates candidate sets, while MADRec performs agent-based reasoning and ranking over the retrieved candidates. To further examine the generalizability of MADRec beyond sequential recommendation, we additionally evaluate it under different recommendation scenarios. The detailed experimental results are reported in Appendix~\ref{app:item_rec}.

\subsubsection{Datasets}
To evaluate the proposed method, we conduct experiments on three real-world datasets: ML-1M\footnote{https://grouplens.org/datasets/movielens/}, Amazon-Beauty, and Amazon-Book\footnote{https://nijianmo.github.io/amazon/}. Following common practice~\cite{sasrec,sun2019bert4rec}, we organize users’ interactions chronologically based on timestamps. Dataset statistics are reported in Table~\ref{Statistical}.
\vspace{-10pt}
\begin{table}[htbp]
    \centering
    \caption{Statistics of the datasets used in our experiments.}
    \vspace{-10pt}
    \renewcommand{\arraystretch}{0.9}
    \setlength\tabcolsep{1.2pt}
    \begin{tabular}{c|cccc}
    \hline
                  & Users & Items & Interactions & Sparsity \\ \hline
    Amazon-Beauty & 8,742 & 5,506 & 88,404 & 99.82\%  \\
    Amazon-Book   & 158,349 & 97,566 & 3,876,663 & 99.97\%  \\
    ML-1M         & 6,040 & 3,884 & 1,000,209 & 95.74\%  \\ 
    \hline
    
    \end{tabular}
    \label{Statistical}
\end{table}
\vspace{-18pt}
\subsubsection{Evaluation Metrics}
Following the leave-one-out evaluation protocol commonly used in sequential recommendation~\cite{sasrec,sun2019bert4rec}, we use the most recent interaction of each user for testing, the second most recent interaction for validation, and the remaining interactions for training. For evaluation, we adopt two widely used top-$K$ ranking metrics: Recall@10 and NDCG@10.

Considering the high inference cost of LLM-based ranking, and following previous studies~\cite{chan2024chateval,llmrank}, we evaluate each method on 2,000 randomly sampled test users per dataset. Sampling protocols of a similar scale have also been widely adopted in recent LLM-based recommendation studies~\cite{zhang2024agentcf,agent4rec,toolrec}. We conduct three runs with independently resampled user subsets, while ensuring that all baselines are evaluated on the same users within each run under an identical protocol. The mean and standard deviation across the three runs are reported in Table~\ref{tab:performance}.

For methods that rank retrieved candidates, users whose
held-out target items are absent from the candidate sets
remain included in the evaluation and receive zero
Recall@10 and NDCG@10.

\subsubsection{Baselines}
We compare MADRec with representative recommendation baselines, including traditional recommendation models, such as SASRec~\cite{sasrec}, S3Rec~\cite{s3rec}, BERT4Rec~\cite{sun2019bert4rec}, and P5~\cite{p5}, as well as LLM-based recommendation methods, including ChatRec~\cite{chatrec}, ToolRec~\cite{toolrec}, LLMRank~\cite{llmrank}, RecMind~\cite{wang2024recmind}, MACF~\cite{macf}, ThinkRec~\cite{thinkrec}, InteRecAgent~\cite{interecagent}, TALLRec~\cite{bao2023tallrec}, MSR-Rec~\cite{wang2026msrec}, and RecPO~\cite{recpo}.

\subsubsection{Implementation Details}
For candidate retrieval, we use SASRec~\cite{sasrec} and
BERT4Rec~\cite{sun2019bert4rec} as retrieval backbones,
yielding MADRec and MADRec$_B$, respectively. The retrieval
models are trained separately from the agent-based ranking
framework and remain frozen during recommendation inference.
Given a user's historical interaction sequence, the retriever
scores items in the item catalog and selects the top-$N$
candidates for subsequent agent-based ranking. Unless
otherwise specified, we set $N=30$. Baselines evaluated
under the shared-candidate reranking protocol receive the
same SASRec Top-30 candidates as the default MADRec.
Other baselines retain their own candidate generation
procedures.

The compared methods use the same available user interaction
histories and item metadata where applicable. MADRec
additionally uses its constructed evidence base as part
of the proposed framework. Recommendation examples used
for in-context learning are drawn exclusively from the
training split. Held-out test interactions and relevance
labels are excluded from model training, prompt
demonstrations, and agent inputs.

MADRec uses Qwen-Flash (qwen-flash-2025-07-28) as its
LLM backbone without updating the LLM parameters.
The same backbone is used for agent-based baselines
whose LLM backbones are interchangeable. Fine-tuned
baselines retain their respective model architectures
and training procedures, with task-specific adaptation
performed on the training split of each dataset.
Unless otherwise specified, the number of debate rounds
is set to $T=2$.

For ECV, we fine-tune a DeBERTaV3-base encoder with a binary
classification head on labeled evidence-claim pairs
constructed from item-level evidence and factual claims
generated by the Semantic Agent. A single verifier is jointly
trained across the three domains and remains frozen during
recommendation inference. We use AdamW with a learning rate
of $2\times10^{-5}$, a batch size of 16, a weight decay of 0.01,
and a maximum input length of 512 tokens. Training runs for
three epochs, and the checkpoint with the highest validation
Macro-F1 is selected. Details of data construction, annotation,
and evaluation are provided in Appendix~\ref{app:ecv}.

\begin{table}[t]
\centering
\caption{Performance comparison on three real-world datasets. Best results are in \textbf{bold}, and the strongest non-MADRec baseline results are \underline{underlined}. Improvements are computed by comparing MADRec with the strongest non-MADRec baseline.}
\vspace{-10pt}
\label{tab:performance}
\setlength{\tabcolsep}{3pt}
\renewcommand{\arraystretch}{0.88}
\small
\begin{adjustbox}{max width=\linewidth}
\begin{tabular}{@{}lcccccc@{}}
\toprule
\multirow{2}{*}{\textbf{Method}}
& \multicolumn{2}{c}{\textbf{ML-1M}}
& \multicolumn{2}{c}{\textbf{Amazon-Book}}
& \multicolumn{2}{c}{\textbf{Amazon-Beauty}} \\
\cmidrule(lr){2-3} \cmidrule(lr){4-5} \cmidrule(lr){6-7}
& Recall@10 & NDCG@10
& Recall@10 & NDCG@10
& Recall@10 & NDCG@10 \\
\midrule

SASRec
& $0.175{\pm0.021}$ & $0.0914{\pm0.009}$
& $0.042{\pm0.013}$ & $0.0186{\pm0.006}$
& $0.072{\pm0.005}$ & $0.0524{\pm0.006}$ \\
BERT4Rec
& $0.138{\pm0.012}$ & $0.0653{\pm0.011}$
& $0.038{\pm0.011}$ & $0.0197{\pm0.003}$
& $0.065{\pm0.011}$ & $0.0497{\pm0.007}$ \\
P5
& $0.181{\pm0.019}$ & $0.0874{\pm0.007}$
& $0.004{\pm0.001}$ & $0.0021{\pm0.001}$
& $0.068{\pm0.009}$ & $0.0501{\pm0.003}$ \\
S3Rec
& $0.170{\pm0.021}$ & $0.0924{\pm0.008}$
& $0.037{\pm0.011}$ & $0.0184{\pm0.004}$
& $0.071{\pm0.010}$ & $0.0512{\pm0.006}$ \\

\midrule

ToolRec
& $0.182{\pm0.012}$ & $0.1157{\pm0.010}$
& $0.050{\pm0.009}$ & $0.0248{\pm0.009}$
& $0.073{\pm0.002}$ & $0.0562{\pm0.009}$ \\
RecMind
& $0.184{\pm0.011}$ & $0.1175{\pm0.012}$
& $0.048{\pm0.003}$ & $0.0243{\pm0.010}$
& $0.071{\pm0.003}$ & $0.0555{\pm0.002}$ \\
InteRecAgent
& $0.190{\pm0.014}$ & $0.1220{\pm0.018}$
& $0.051{\pm0.011}$ & $0.0251{\pm0.012}$
& $0.075{\pm0.004}$ & $0.0550{\pm0.003}$ \\
ChatRec
& $0.192{\pm0.012}$ & $0.1252{\pm0.016}$
& $0.053{\pm0.010}$ & $0.0253{\pm0.011}$
& $0.078{\pm0.002}$ & $0.0565{\pm0.005}$ \\
LLMRank
& $0.198{\pm0.015}$ & $0.1235{\pm0.014}$
& $0.055{\pm0.007}$ & $0.0256{\pm0.010}$
& $0.081{\pm0.005}$ & $0.0572{\pm0.004}$ \\
ThinkRec
& $0.205{\pm0.014}$ & $0.1260{\pm0.012}$
& $0.057{\pm0.009}$ & $0.0254{\pm0.008}$
& $0.084{\pm0.007}$ & $0.0560{\pm0.002}$ \\
MACF
& $0.206{\pm0.011}$ & $0.1301{\pm0.011}$
& $0.059{\pm0.006}$ & $0.0257{\pm0.014}$
& $0.090{\pm0.009}$ & $0.0581{\pm0.007}$ \\
MSR-Rec 
& $0.210{\pm0.008}$ & $0.1315{\pm0.019}$
& $\underline{0.061}{\pm0.003}$ & $\underline{0.0263}{\pm0.013}$
& $0.093{\pm0.005}$ & $0.0652{\pm0.006}$ \\
TALLRec
& $0.207{\pm0.009}$ & $0.1320{\pm0.015}$
& $0.058{\pm0.007}$ & $0.0250{\pm0.010}$
& $0.098{\pm0.003}$ & $0.0660{\pm0.004}$ \\
RecPO
& $\underline{0.212}{\pm0.010}$ & $\underline{0.1348}{\pm0.014}$
& $0.060{\pm0.006}$ & $0.0253{\pm0.011}$
& $\underline{0.101}{\pm0.004}$ & $\underline{0.0671}{\pm0.005}$ \\
\midrule

MADRec
& $0.242{\pm0.011}$ & $0.1487{\pm0.011}$
& $0.081{\pm0.004}$ & $0.0324{\pm0.008}$
& $\mathbf{0.124}{\pm0.003}$ & $\mathbf{0.0760}{\pm0.001}$ \\
MADRec$_B$
& $0.231{\pm0.012}$ & $0.1372{\pm0.008}$
& $0.070{\pm0.002}$ & $0.0301{\pm0.009}$
& $0.114{\pm0.002}$ & $0.0692{\pm0.002}$ \\
MADRec\textsubscript{DeepSeek}
& ${0.238}{\pm0.015}$ & ${0.1482}{\pm0.013}$
& ${0.077}{\pm0.003}$ & ${0.0315}{\pm0.010}$
& ${0.120}{\pm0.005}$ & ${0.0753}{\pm0.001}$ \\
MADRec\textsubscript{GPT}
& $\mathbf{0.249}{\pm0.013}$ & $\mathbf{0.1504}{\pm0.009}$
& $\mathbf{0.090}{\pm0.006}$ & $\mathbf{0.0341}{\pm0.005}$
& $0.121{\pm0.004}$ & $0.0758{\pm0.004}$ \\
MADRec\textsubscript{LLaMA}
& $0.240{\pm0.010}$ & $0.1470{\pm0.011}$
& $0.076{\pm0.009}$ & $0.0317{\pm0.014}$
& $0.118{\pm0.007}$ & $0.0745{\pm0.003}$ \\

\midrule
\textbf{Improvement}
& $\mathbf{14.15\%}$ & $\mathbf{10.31\%}$
& $\mathbf{32.79\%}$ & $\mathbf{23.19\%}$
& $\mathbf{22.77\%}$ & $\mathbf{13.26\%}$ \\
\bottomrule
\end{tabular}
\end{adjustbox}
\vspace{-10pt}
\end{table}
\subsection{Performance Comparison}
Table~\ref{tab:performance} summarizes the performance of various recommendation methods across all three datasets. Overall, the MADRec variants achieve the strongest performance across all three datasets in terms of Recall@10 and NDCG@10, while the default MADRec consistently outperforms all non-MADRec baselines. Its improvements over traditional recommendation models suggest that explicit reasoning over interaction signals and item
semantics can complement latent ID-based preference modeling, enabling more comprehensive and reliable recommendation. More importantly, MADRec consistently outperforms the evaluated LLM-based
recommendation methods, suggesting that effective LLM reasoning alone is insufficient when the reliability of intermediate judgments is not explicitly examined during agent collaboration. MADRec addresses this
limitation through cross-agent debate that exposes potentially inconsistent judgments, item-level evidence that constrains unsupported semantic inferences, and reliability-aware calibration that reduces the influence of unreliable outputs on the final ranking.

In addition, MADRec and MADRec$_B$ consistently outperform their corresponding retrieval backbones, SASRec and BERT4Rec, respectively. Since each MADRec variant uses the same pretrained and frozen retrieval backbone as its backbone-only counterpart, these paired comparisons demonstrate the effectiveness of the subsequent agent-based ranking stage. The consistent improvements across SASRec and BERT4Rec further indicate that MADRec can be applied to different retrieval architectures. We also evaluate MADRec with different LLM backbones by replacing Qwen-Flash with DeepSeek-v3, GPT-4o, and LLaMA-3-8B-Instruct. MADRec achieves consistently strong performance across these LLMs, indicating that its effectiveness is not tied to a particular retrieval architecture or language model.

\subsection{Ablation Study}

To examine the contributions of individual modules and the role
of cross-agent interaction, we conduct component ablations and
controlled comparisons of alternative reasoning strategies.
All variants use the same evaluation users, retrieval backbone,
candidate sets, and LLM backbone as MADRec.

\begin{figure}[!ht]
    \centering
    \includegraphics[width=0.9\linewidth]{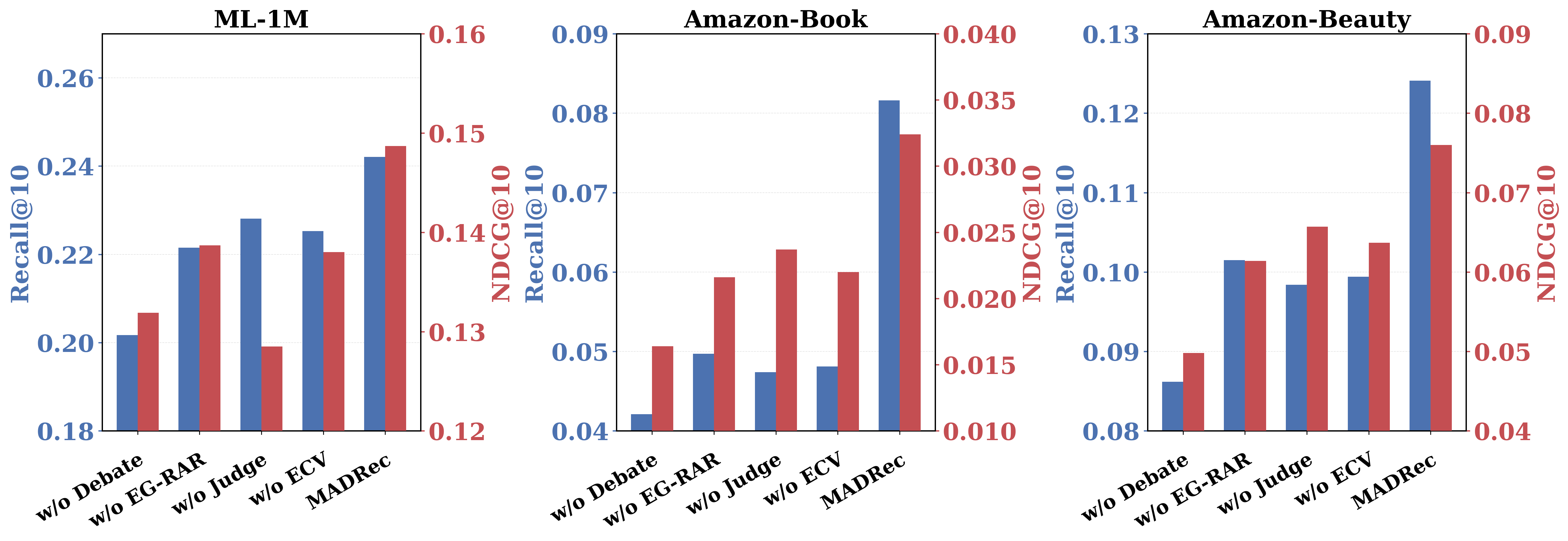}
    \vspace{-10pt}
    \caption{Ablation study of MADRec on three datasets.}
    \vspace{-15pt}
    \label{fig_ablation}
\end{figure}

\subsubsection{Component Ablations.}
We consider the following variants:
\begin{itemize}
    \item \textbf{w/o EG-RAR}: removes the evidence-grounded
    reasoning module, including the retrieved item-level evidence,
    the Claim-Evidence Table, and evidence consistency validation.

    \item \textbf{w/o ECV}: removes the evidence consistency
    validator while retaining the same item-level evidence,
    Claim-Evidence Table, and debate process. The semantic score
    is computed as $p_{S,i}^{(t)}\rho_{S,i}^{(t)}$,
    without adjustment based on evidence support.

    \item \textbf{w/o Debate}: removes cross-agent feedback.
    The Interaction Agent and Semantic Agent independently
    evaluate the same candidates, and their outputs are passed
    to the Judge Agent for aggregation.

    \item \textbf{w/o Judge}: replaces the Judge's adaptive
    weighting with fixed equal weights,
    $\omega_I=\omega_S=0.5$, while preserving the final candidate
    union, branch scores, and score-based ranking procedure.
\end{itemize}

Figure~\ref{fig_ablation} presents the component ablation results. Removing EG-RAR reduces recommendation performance, suggesting that item-level evidence grounding contributes to candidate evaluation. The comparison between MADRec and w/o ECV further demonstrates the importance of independent evidence verification. Although both variants retain the same evidence and claim–evidence organization, w/o ECV lacks explicit verification of semantic claims. In contrast, MADRec uses evidence support probabilities to adjust semantic scores. This helps reduce the influence of unsupported semantic claims on candidate rankings and improves recommendation performance.
\vspace{-10pt}
\begin{figure}[!ht]
    \centering
    \includegraphics[width=0.9\linewidth]{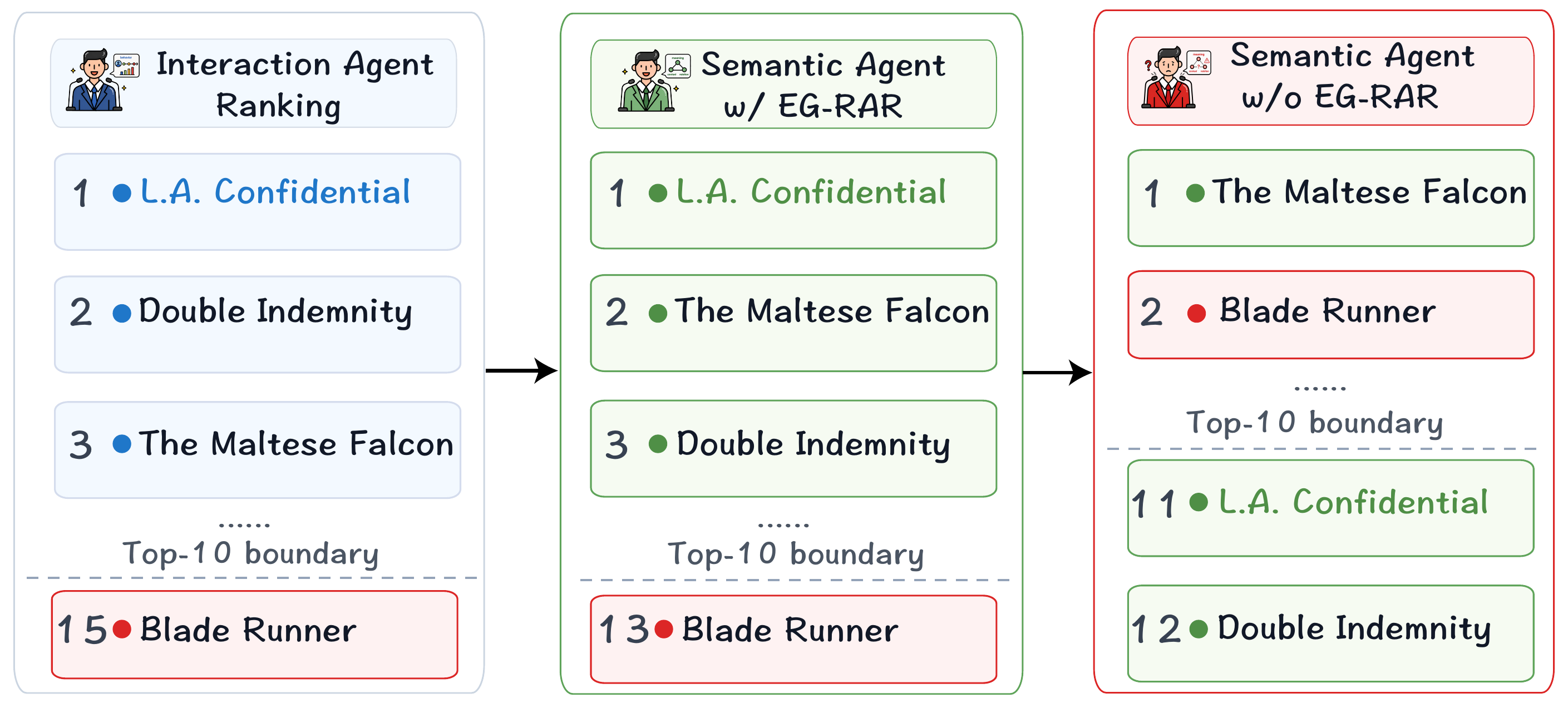}
    \vspace{-10pt}
    \caption{EG-RAR promotes evidence-supported candidates and suppresses weakly grounded ones.}
    \vspace{-10pt}
    \label{ablation_EG-RAR}
\end{figure}
A representative case illustrates the role of EG-RAR.
As shown in Figure~\ref{ablation_EG-RAR}, the Interaction Agent
ranks the ground-truth item \textit{L.A. Confidential} first,
consistent with the user's preference for crime and film-noir
films. After observing this ranking, the Semantic Agent evaluates
the candidates from a content-based perspective. With EG-RAR,
it retains \textit{L.A. Confidential} at Rank~1 and keeps
\textit{Blade Runner} outside the Top-10. Without EG-RAR,
it promotes \textit{Blade Runner} to Rank~2 and demotes
\textit{L.A. Confidential} to Rank~11. This example illustrates
how evidence-grounded semantic evaluation can affect candidate
rankings.

Removing Debate reduces recommendation performance, as the two agents independently assess candidates without opportunities to challenge or revise each other's judgments. This suggests that complementary interaction and semantic signals alone are insufficient; their iterative exchange helps identify inconsistent candidate assessments and refine the resulting rankings. We further investigate the role of reciprocal feedback through subsequent comparisons with self-refinement and one-way interaction under the same LLM call budgets.

Finally, replacing the Judge's adaptive weights with equal weights
reduces recommendation performance under the same final
candidate set and branch scores. This suggests that
context-dependent weighting improves the fusion of
interaction-based and semantic scores. The subsequent
weight analysis on ML-1M examines whether the Judge
assigns more weight to the branch with higher per-user
NDCG@10.
    \vspace{-10pt}
\begin{figure}[!ht]
    \centering
    \includegraphics[width=0.9\linewidth]{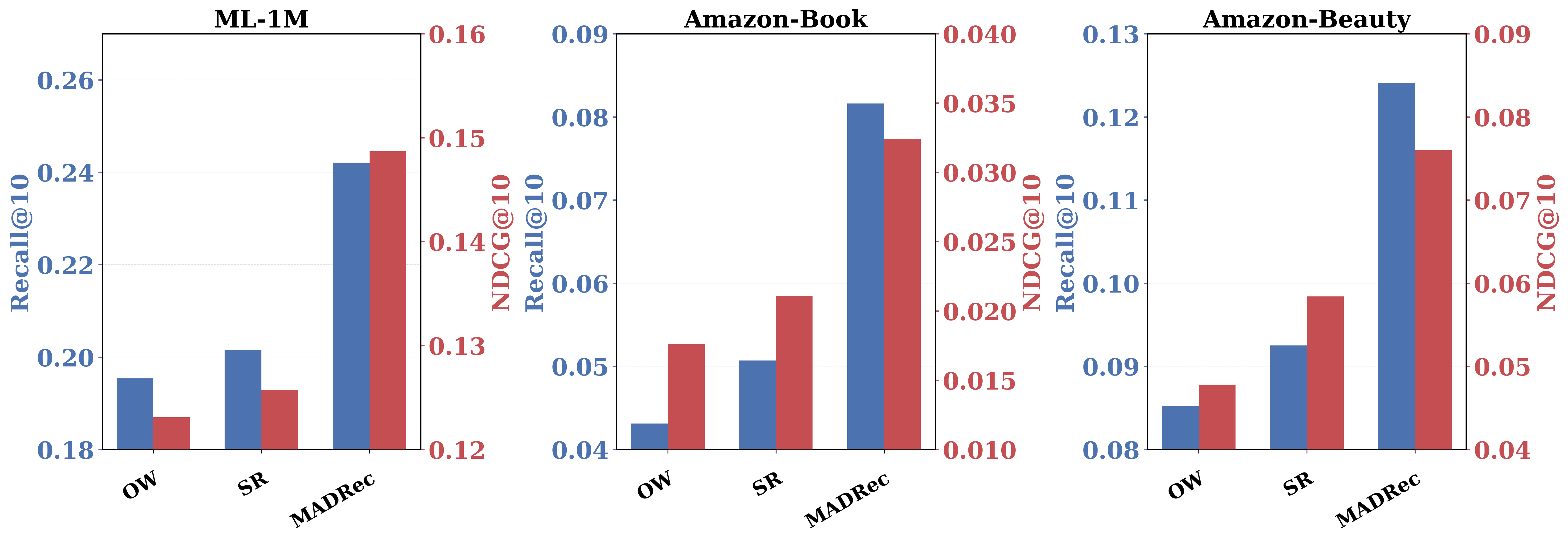}
    \vspace{-10pt}
    \caption{Interaction ablation study of MADRec on three datasets.
OW and SR denote one-way interaction and self-refinement}
    \vspace{-15pt}
    \label{interaction_ablation}
\end{figure}

\subsubsection{Comparison of Interaction Strategies.}
We compare MADRec with two additional variants to examine
whether its gains arise from repeated inference and whether
reciprocal feedback provides additional value.

\begin{itemize}
    \item \textbf{Self-refinement}: A single agent evaluates candidates
    using both interaction signals and item semantics. It receives
    the same inputs as MADRec and uses the same frozen ECV and
    semantic score adjustment. With the same number of LLM calls
    as MADRec, the agent generates initial judgments and then
    iteratively reviews and revises them. 

    \item \textbf{One-way interaction}: This variant retains both
    agents, EG-RAR, and the final Judge. Feedback flows only
    from the Interaction Agent to the Semantic Agent within
    each round. In later rounds, the Interaction Agent revises
    its judgments using the user context and its own previous
    outputs. The variant uses the same number of rounds and
    LLM calls as MADRec.
\end{itemize}

As shown in Figure~\ref{interaction_ablation}, MADRec
consistently outperforms both Self-refinement and One-way
interaction in terms of Recall@10
and NDCG@10. Its advantage over Self-refinement under the
same LLM call budget suggests that the performance gains
cannot be explained solely by additional inference calls.
Exchanging judgments between agents with complementary
perspectives provides benefits beyond repeatedly revising
a single agent's own outputs.

MADRec also achieves better performance than One-way
interaction, supporting the value of reciprocal feedback.
While one-way interaction allows the Semantic Agent to
consider interaction-based judgments, it prevents the
Interaction Agent from revisiting its decisions in response
to semantic feedback. The results suggest that enabling
both agents to revise their judgments through reciprocal
exchanges improves the final recommendation ranking.

\subsubsection{Analysis of Judge Weights}
To examine whether the Judge's adaptive weights reflect
the relative ranking quality of the two branches, we
analyze 2,000 test users on the ML-1M. For each user, we use the Interaction Agent's
final-round ranking and the Semantic Agent's ECV-adjusted
final-round ranking, both obtained before Judge aggregation. After inference, we compute NDCG@10 for both rankings using the ground-truth items. These scores serve as a reference for evaluating the Judge's weight assignments.

We examine whether the Judge assigns more weight to the branch
with the higher NDCG@10. Of the 2,000 test users, 420 had
different NDCG@10 scores for the two branches. We evaluate
agreement on these users and count equal weights as incorrect.
The Judge assigned more weight to the better-performing branch
for 384 of the 420 users (91.4\%).

We use ML-1M as a case study to further analyze whether the Judge's adaptive weights reflect the relative ranking quality of the two branches. Among the 420 users for whom the two branches achieved different NDCG@10 scores, we compute two differences by subtracting the semantic branch’s NDCG@10 and weight from the interaction branch’s NDCG@10 and weight, respectively. We then assign each user two ranks. The user with the largest NDCG@10 difference ranks first by performance, and the user with the largest weight difference ($\omega_I-\omega_S$) ranks first by assigned weight. We use Spearman’s rank correlation to measure how closely the two rankings agree, with a value of 1 indicating identical rankings. The observed correlation of 0.89 shows that when the interaction branch has a larger NDCG@10 advantage for one user than for another, the Judge also tends to give it a larger share of the weight for that user. This result further supports the effectiveness of JRC in assigning weights based on the relative ranking quality of the two branches.
\subsection{Case Study}
\begin{figure}[t]
    \centering
    \includegraphics[width=\linewidth]{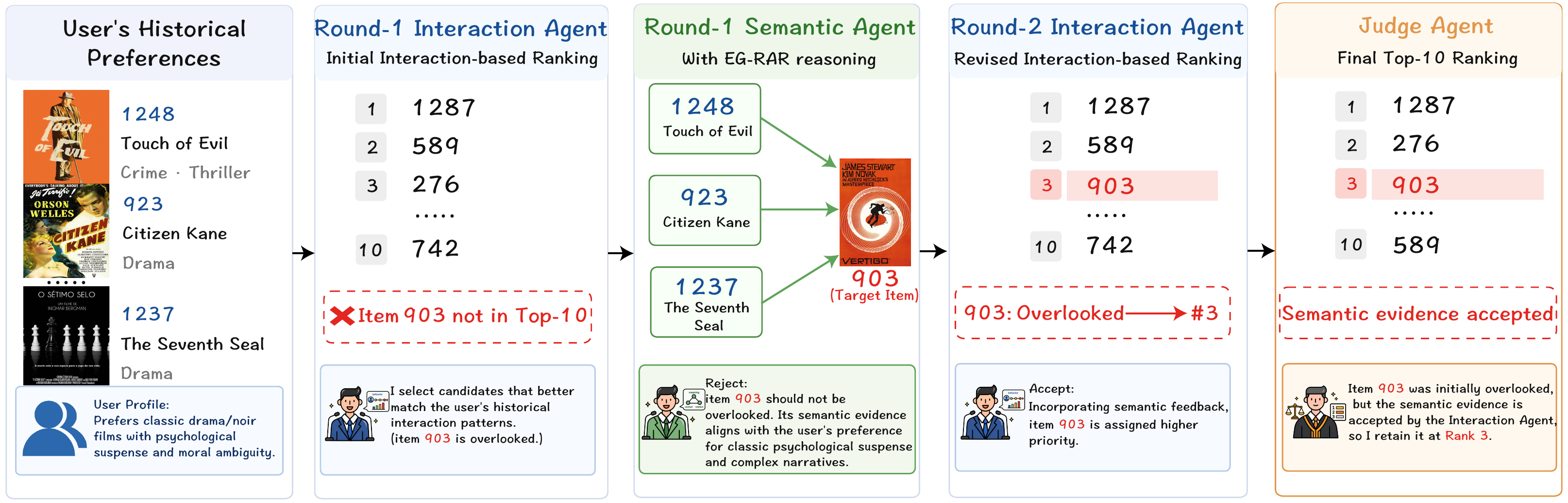}
    \vspace{-20pt}
    \caption{A representative case illustrating how MADRec
    corrects an initial ranking error through multi-agent
    debate, evidence-grounded semantic reasoning, and
    judge-based ranking calibration. The ground-truth
    item (903, \textit{Vertigo}) is initially overlooked
    by the Interaction Agent but is subsequently identified
    through semantic feedback and ranked third in the
    final recommendation.}
    \vspace{-10pt}
    \label{fig:case_study}
\end{figure}
As shown in Figure~\ref{fig:case_study}, we analyze a representative case where the ground-truth item is item 903, namely \textit{Vertigo}. The target item is included in the retrieved candidate set, but it is overlooked by the \textit{Interaction Agent} in the initial Top-10 ranking due to limited direct interaction-based data.

In the first debate round, the \textit{Interaction Agent} mainly relies on historical interaction patterns and collaborative signals, resulting in item 903 being excluded from its initial ranking. The \textit{Semantic Agent} then reexamines the candidate set with \textbf{EG-RAR} and identifies that item 903 has been overlooked. Based on the retrieved item-level evidence, item 903 aligns with the user's latent preference for classic films involving psychological suspense, misperception, observation, and complex narratives, as reflected by historical items such as 1248, 923, and 1237. In the second debate round, the \textit{Interaction Agent} incorporates this semantic feedback and promotes item 903 in its ranking output. Finally, the \textit{Judge Agent} integrates this refined judgment through global ranking calibration and promotes item 903 to the third position.

This case demonstrates how MADRec corrects an initial ranking error through the coordinated effects of its key components: \textbf{MAD} exposes and communicates inconsistent judgments across complementary reasoning perspectives, \textbf{EG-RAR} guides the Semantic Agent toward evidence-grounded and more reliable reasoning, and \textbf{JRC} calibrates and consolidates the resulting heterogeneous judgments into a reliable final ranking, thereby preventing the initial error from propagating to the final decision.

\begin{table}[t]
\centering

\caption{Intermediate ranking error propagation and correction
under controlled interventions (\%). Lower Transfer and
Persistence and higher Correction indicate better performance.}
\vspace{-7pt}
\label{tab:ranking_error_propagation}
\begin{tabular}{@{}lccc@{}}
\toprule
Method & Transfer $\downarrow$ & Correction $\uparrow$
& Persistence $\downarrow$ \\
\midrule
MACRec              & 43.6 & 31.2 & 30.0 \\
MACF                & 35.1 & 45.2 & 19.2 \\
Self-refinement     & 23.6 & 51.2 & 11.5 \\
One-way interaction & 25.1 & 55.2 & 11.2 \\
MADRec              & \textbf{14.7} & \textbf{87.9} & \textbf{1.8} \\
\bottomrule
\vspace{-20pt}
\end{tabular}
\end{table}

\subsection{Reliability and Robustness Analysis}
\subsubsection{Intermediate Ranking Error Propagation}

To assess MADRec's ability to limit the propagation of intermediate
ranking errors and correct them in later stages, we compare MADRec
with the baselines under controlled ranking perturbations.

We apply the intervention only when the target item appears in the intermediate Top-10. Before passing the list to the next reasoning stage, we replace the target with an item from positions 21--30
in the original ranking. This change affects only the list sent
to the next stage. The target remains in the candidate set, and
the candidate scores and ranking rationales remain unchanged. We denote the set of eligible interventions by $\mathcal{I}$.
\emph{Transfer} is the fraction of these interventions in which
the target item falls outside the Top-10 at the next stage.
These cases form the set $\mathcal{T}$.
\emph{Correction} is the fraction of cases in $\mathcal{T}$ in
which the target returns to the final Top-10. We denote these
corrected cases by $\mathcal{C}$.
\emph{Persistence} is the fraction of all eligible interventions
in which the target is absent from both the next-stage Top-10
and the final Top-10:
\begin{equation}
\label{eq:error_propagation}
\mathrm{Transfer} = \frac{|\mathcal{T}|}{|\mathcal{I}|},
\quad
\mathrm{Correction} = \frac{|\mathcal{C}|}{|\mathcal{T}|},
\quad
\mathrm{Persistence} =
\frac{|\mathcal{T}| - |\mathcal{C}|}{|\mathcal{I}|}.
\end{equation}

For MADRec and One-way interaction, we perturb the
Interaction Agent's first-round Top-10 before passing
it to the Semantic Agent. For Self-refinement, we perturb
the initial Top-10 before passing it to the first refinement
step. For MACRec, we intervene at the Reflector-to-Manager
interface. For MACF, we perturb the ranking sent from a
selected user or item agent to the Orchestrator. Each method
then completes its remaining reasoning steps.

As shown in Table~\ref{tab:ranking_error_propagation}, MADRec achieves the lowest Transfer and Persistence rates, while obtaining the highest Correction rate among the evaluated methods. The comparison with Self-refinement shows that
MADRec's multi-agent reasoning process is more robust to
intermediate ranking errors than repeated revision by a
single agent. Compared with One-way interaction, MADRec
recovers transferred errors more often. This pattern is
consistent with reciprocal feedback giving agents more
opportunities to revisit earlier ranking judgments and
restore affected target items before final aggregation. MADRec also outperforms MACRec and MACF under the same intervention protocol. This advantage comes from its multi-agent debate and item-level evidence, which allow agents to revisit earlier ranking decisions and reassess candidate items based on supporting evidence.
\vspace{-10pt}
\begin{figure}[htbp]
    \centering
    \includegraphics[width=0.8\linewidth]{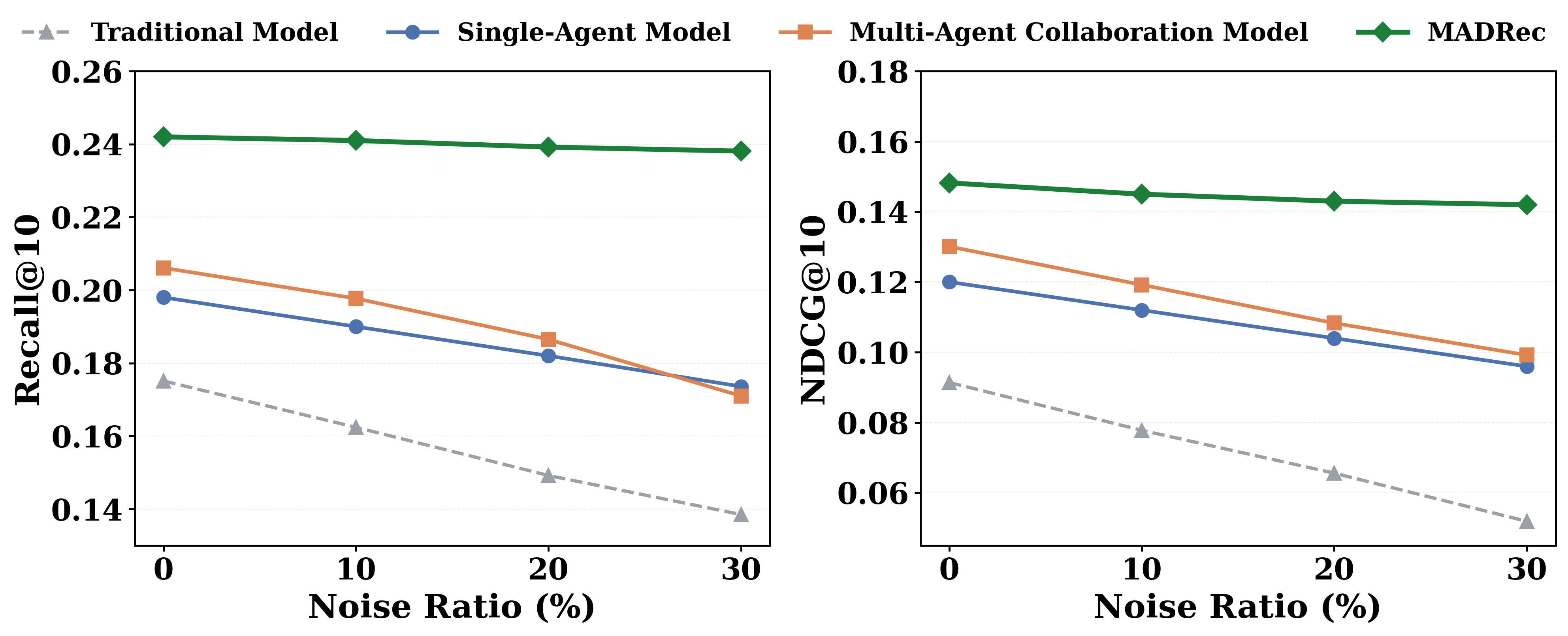}
    \Description{Recall at 10 and NDCG at 10 decline as interaction noise increases. MADRec stays above SASRec, DRDT, and MACF at the tested noise levels.}
    \vspace{-5pt}
    \caption{Effect of dataset noise on R@10 and N@10 across traditional model, single-agent model, multi-agent model, and MADRec.}
    \vspace{-20pt}
    \label{fig_propagation}
\end{figure}
\subsubsection{Robustness to Noisy Interactions}
To assess robustness to corrupted behavioral inputs, we perturb users' historical interaction sequences by randomly selecting a proportion of positions and replacing their items with items sampled from the item pool. The held-out ground-truth items remain unchanged. We increase the noise ratio from 0\% to 30\% and evaluate recommendation performance using Recall@10 and NDCG@10.

Figure~\ref{fig_propagation} compares MADRec with SASRec~\cite{sasrec}, a conventional sequential recommender; DRDT~\cite{wang2023drdt}, a single-agent reasoning method; and MACF~\cite{macf}, a multi-agent recommendation framework. As the noise ratio increases, all methods exhibit performance degradation, showing that corrupted interaction histories adversely affect recommendation quality. Nevertheless, MADRec maintains higher Recall@10 and NDCG@10 than the compared baselines across the evaluated noise levels. These results complement the inter-agent intervention analysis by assessing the robustness of the overall framework to corrupted behavioral inputs.

\subsubsection{Evidence Verification and Ranking Effectiveness}

To evaluate ECV's effectiveness in assessing claim-evidence
support and its contribution to recommendation quality,
we conduct two complementary evaluations. First, we compare
ECV with LLM self-verification on evidence-claim
classification. Second, we replace ECV in MADRec with
LLM self-verification and compare its recommendation
performance with that of full MADRec.

For LLM self-verification, we use the same LLM to generate and verify claims. ECV and LLM self-verification use the same input: a claim and its corresponding evidence passage. Both classify the claim as $\mathrm{Supported}$ or $\mathrm{NotSupported}$ according to the annotation criteria in Appendix~\ref{app:ecv}. We evaluate their predictions
against the same reference annotations and report Macro-F1, class-wise precision and recall across the three datasets. For the recommendation comparison, we retain identical
evaluation inputs and all other framework settings,
including the semantic score adjustment formula.
Full MADRec uses ECV's predicted probability of
$\mathrm{Supported}$ as the support coefficient,
whereas the self-verification variant assigns a value of
1 to $\mathrm{Supported}$ and 0 to $\mathrm{NotSupported}$.

As shown in Table~\ref{tab:ecv_verification}, ECV achieves
higher Macro-F1 and better class-wise precision and recall
than LLM self-verification across all three datasets.
These results show that ECV is better at identifying whether a claim is supported by the provided evidence. This capability enables ECV to provide
an independent assessment of semantic claims beyond
the Semantic Agent's own judgments, helping verify whether intermediate recommendation reasoning is supported by evidence. Full MADRec also achieves higher Recall@10 and NDCG@10
than its self-verification variant. Within MADRec, ECV predicts how well the evidence supports each semantic claim. It then uses these probabilities to adjust the scores of semantic judgments before ranking. As a result, semantic judgments with less evidence support receive lower scores. These adjusted scores are then used in subsequent agent interactions and final ranking aggregation. The recommendation results support the
downstream effectiveness of this verification and score adjustment scheme.
\begin{table}[t]
\centering
\caption{Evidence verification and downstream recommendation
performance. $P_c$ and $R_c$ denote precision and recall for
class $c$, with $\mathrm{S}$ and $\mathrm{NS}$ denoting
\texttt{Supported} and \texttt{NotSupported}, respectively.}
\vspace{-10pt}
\label{tab:ecv_verification}
\scriptsize
\setlength{\tabcolsep}{3pt}
\begin{adjustbox}{max width=\linewidth}
\begin{tabular}{@{}llccccccc@{}}
\toprule
& & \multicolumn{5}{c}{Evidence verification (\%)}
  & \multicolumn{2}{c}{Recommendation} \\
\cmidrule(lr){3-7}\cmidrule(l){8-9}
Dataset & Variant
& Macro-F1 $\uparrow$
& $P_{\mathrm{NS}}\uparrow$
& $R_{\mathrm{NS}}\uparrow$
& $P_{\mathrm{S}}\uparrow$
& $R_{\mathrm{S}}\uparrow$
& R@10 $\uparrow$
& N@10 $\uparrow$ \\
\midrule
ML-1M
& LLM self-verification
& 82.73 & 74.47 & 70.00 & 92.61 & 94.00 & 0.235 & 0.1429 \\
& MADRec (w/ ECV)
& $\mathbf{86.78}$
& $\mathbf{79.59}$
& $\mathbf{78.00}$
& $\mathbf{94.53}$
& $\mathbf{95.00}$
& \textbf{0.242} & \textbf{0.1487} \\
\midrule
Amazon-Book
& LLM self-verification
& 81.48 & 75.70 & 67.50 & 90.08 & 93.16 & 0.077 & 0.0303 \\
& MADRec (w/ ECV)
& $\mathbf{85.17}$
& $\mathbf{79.65}$
& $\mathbf{75.00}$
& $\mathbf{92.25}$
& $\mathbf{93.95}$
& \textbf{0.081} & \textbf{0.0324} \\
\midrule
Amazon-Beauty
& LLM self-verification
& 81.91 & 72.62 & 67.78 & 93.03 & 94.39 & 0.119 & 0.0723 \\
& MADRec (w/ ECV)
& $\mathbf{85.65}$
& $\mathbf{77.27}$
& $\mathbf{75.56}$
& $\mathbf{94.66}$
& $\mathbf{95.12}$
& \textbf{0.124} & \textbf{0.0760} \\
\bottomrule
\vspace{-20pt}
\end{tabular}
\end{adjustbox}
\end{table}

\subsection{Parameter Sensitivity Analysis}
\begin{figure}[htbp]
    \centering
    \includegraphics[width=0.75\linewidth]{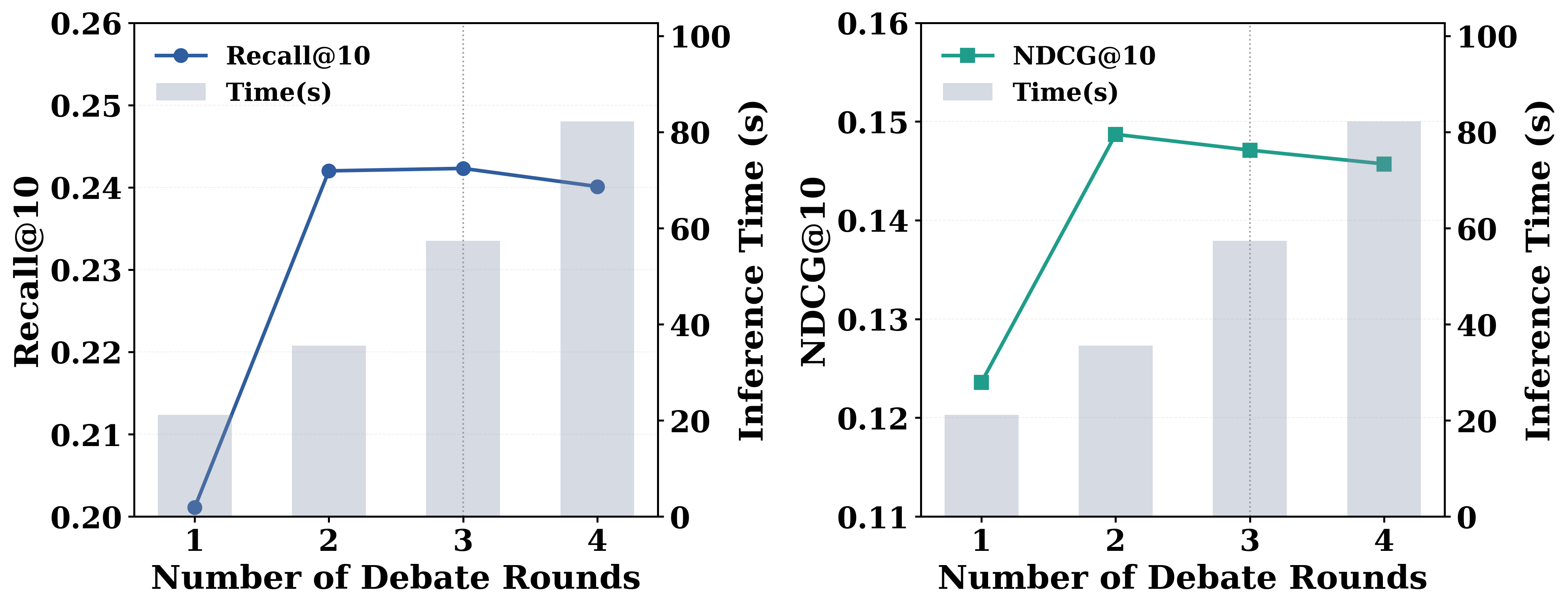}
    \Description{Recall at 10 and NDCG at 10 on ML-1M improve from one to two debate rounds, with no further gain from additional rounds.}
    \vspace{-10pt}
    \caption{Effect of debate rounds.}
    \vspace{-10pt}
    \label{fig_debate}
\end{figure}

We analyze the sensitivity of MADRec to the number of debate rounds on the ML-1M dataset. As shown in Figure~\ref{fig_debate}, using only one debate round yields lower performance because the agents have limited opportunities to refine their judgments through cross-agent feedback. Increasing the number of rounds to two clearly improves performance, indicating that an additional round helps identify and reassess inconsistent judgments between interaction-based and semantic perspectives. However, further increasing the number of rounds provides no additional gains and introduces redundant reasoning and higher inference cost. Therefore, we set the number of debate rounds to $T=2$.
\vspace{-10pt}
\vspace{-5pt}

\begin{figure}[htbp]
    \centering
    \includegraphics[width=0.75\linewidth]{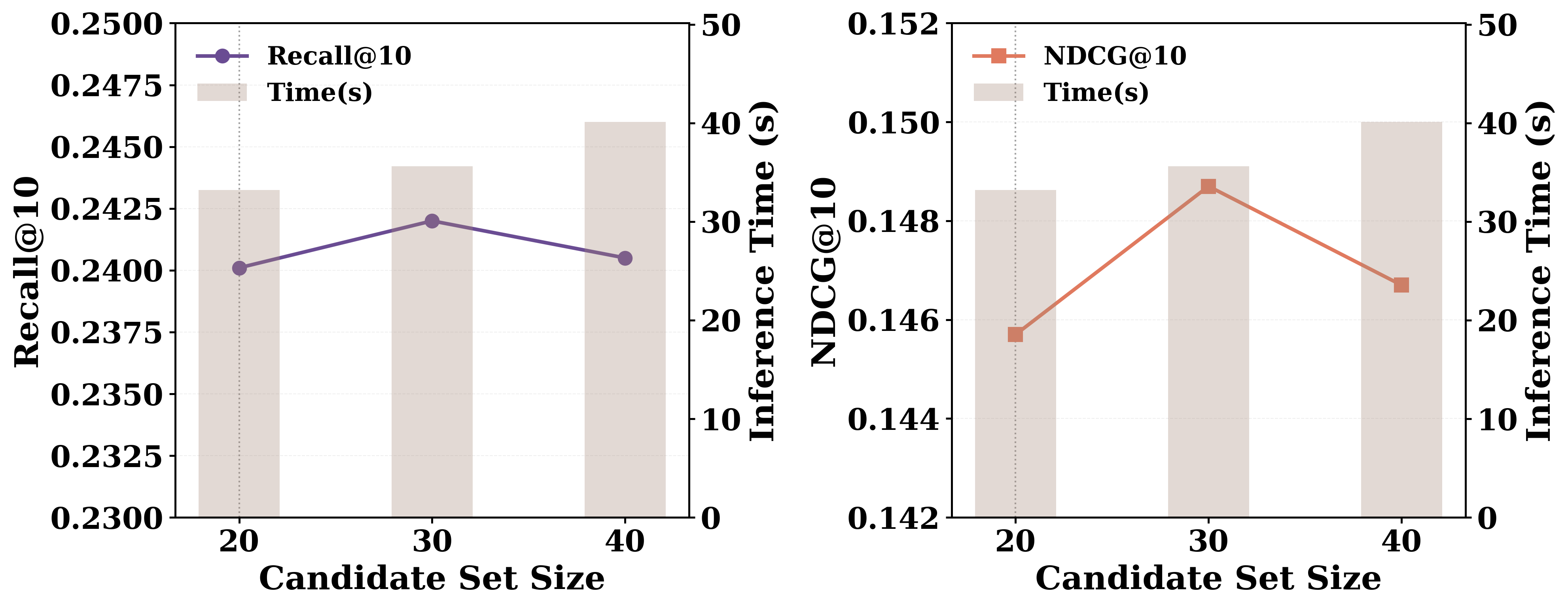}
    \Description{On ML-1M, performance improves when the candidate pool grows from 20 to 30 items and levels off at 40.}
    \vspace{-10pt}
    \caption{Effect of candidate set size.}
    \vspace{-5pt}
    \label{fig_candidate}
\end{figure}

We further evaluate the sensitivity of MADRec to the candidate set size on the ML-1M dataset by varying it among 20, 30, and 40. As shown in Figure~\ref{fig_candidate}, increasing the candidate set size from 20 to 30 improves performance, suggesting that a moderately larger candidate set provides more potentially relevant items for fine-grained ranking. When the size further increases to 40, performance tends to saturate, while inference cost continues to increase. This indicates that additional candidates may introduce redundant or less informative items without providing useful preference signals. Considering both effectiveness and efficiency, we set the candidate set size to 30.
\vspace{-10pt}

\subsection{Efficiency Analysis}

To assess the computational cost of MADRec and its trade-off
with recommendation quality, we compare it with representative
tool- and agent-based baselines. We report NDCG@10 alongside the average number of LLM calls, input and output tokens, end-to-end latency, and API cost per user. 

Measurements are conducted on the same 2000 test users from ML-1M, with requests processed sequentially. End-to-end latency is measured from receiving the user input to producing the final ranking. For MADRec, this includes candidate retrieval, online evidence retrieval, agent inference, ECV evaluation, Judge-based aggregation, output parsing, and any retries.

As shown in Table~\ref{tab:efficiency}, MADRec achieves the highest NDCG@10 among the compared methods while maintaining competitive inference efficiency with relatively few LLM calls. Although InteRecAgent incurs lower end-to-end latency and per-user API cost, its recommendation performance is substantially inferior to that of MADRec. This method plans tool calls ahead of time and uses specialized recommendation tools to retrieve and rank items which helps explain its efficiency. In contrast, MADRec improves candidate ranking through multi-round cross-agent interactions and evidence support verification, achieving a substantial gain in NDCG@10 at a relatively modest additional latency and monetary cost. These results demonstrate a favorable trade-off between recommendation quality and inference overhead. Future work will focus on streamlining agent interactions and reducing contextual input requirements to further lower online inference costs while preserving recommendation effectiveness.

%% file: sect5_conclution.tex
\begin{table}[t]
\centering
\caption{Recommendation performance and inference cost on ML-1M.}
\vspace{-5pt}
\label{tab:efficiency}
\small
\setlength{\tabcolsep}{5pt}
\begin{adjustbox}{max width=\linewidth}
\begin{tabular}{lccccc}
\toprule
Method
& NDCG@10 $\uparrow$
& LLM Calls $\downarrow$
& \makecell{Input/Output\\Tokens (K) $\downarrow$}
& \makecell{Latency\\(s/user) $\downarrow$}
& \makecell{API Cost\\(\$/user) $\downarrow$} \\
\midrule
InteRecAgent
& $0.1220{\pm0.018}$
& 4
& 32.0 / 8.0
& 34.2
& 0.0048 \\

MACF
& $0.1301{\pm0.011}$
& 8
& 80.0 / 13.0
& 62.3
& 0.0092 \\

ToolRec
& $0.1157{\pm0.010}$
& 10
& 34.7 / 10.9
& 57.4
& 0.0061 \\

MADRec
& $0.1487{\pm0.011}$
& 5
& 36.0 / 8.3
& 35.6
& 0.0051 \\
\bottomrule
\vspace{-25pt}
\end{tabular}
\end{adjustbox}

\end{table}
\section{Conclusion}
In this paper, we propose MADRec, an evidence-grounded multi-agent recommendation framework designed to mitigate unreliable intermediate judgments and their propagation during recommendation reasoning. Specifically, \textbf{MAD} enables reciprocal, multi-round interactions between the Interaction Agent and Semantic Agent, allowing candidate-level ranking 
decisions to be iteratively examined and refined from complementary perspectives. \textbf{EG-RAR} incorporates item-level evidence into semantic reasoning and employs an independent evidence consistency validator to assess claim–evidence support, thereby reducing the influence of insufficiently grounded semantic judgments. \textbf{JRC} further integrates the complementary agent outputs through reliability-aware ranking calibration to produce the final recommendations.

Extensive experiments on three real-world datasets demonstrate the effectiveness of MADRec across different retrieval and LLM backbones. Controlled ablation studies highlight the contribution of cross-agent debate, while error propagation experiments show that MADRec reduces the transfer of injected ranking errors and facilitates their subsequent correction. Robustness analyses further demonstrate its ability to maintain recommendation quality under noisy interaction histories. Despite these improvements, multi-round LLM interactions introduce non-negligible inference latency and computational costs, limiting practical deployment in latency-sensitive scenarios. Future work will investigate more efficient and adaptive debate strategies and extend MADRec to recommendation settings involving richer multimodal and contextual information.

%% file: sect6_acknowledgments.tex

%% file: sect7_appendix.tex
\section{Prompt Details}
\label{app:details}

\begin{tcolorbox}[
    title={Prompt Template for the Interaction Agent},
    colback=gray!5,
    colframe=gray!60,
    breakable,
    fonttitle=\bfseries
]
\small
\ttfamily

"You are the Interaction Preference Expert.\textbackslash n\textbackslash n%
Scope:\textbackslash n%
Produce an independent evaluation and ranking based on the user's
interaction history. Use feedback from the Semantic Agent to
re-examine interaction-based assumptions and identify preference
signals that may have been overlooked.\textbackslash n\textbackslash n%
Task:\textbackslash n%
Evaluate every eligible item in the candidate pool using historical
interaction signals, including frequency, recency, and rating
patterns. Produce an interaction-based rationale, preference score,
and critique for each item. Select the Top-10 items with the highest
interaction-based preference scores.\textbackslash n\textbackslash n%
Instructions:\textbackslash n\textbackslash n%
Follow three steps: Observation, Thought, and Action.
\textbackslash n\textbackslash n%
* Observation:\textbackslash n%
In the first round, begin with the supplied user information and
candidate pool. When semantic feedback is available, review the
processed candidate records and semantic Top-10 ranking from
previous rounds. These records contain the factual claim, quoted
evidence, preference score, confidence score, critique, ECV support
coefficient, and adjusted semantic score. Use the supplied ECV
support coefficient to gauge how strongly the paired evidence
supports the factual claim. Use the adjusted semantic score to
interpret the semantic ranking. Identify agreements and disagreements
with your interaction-based assessment.\textbackslash n\textbackslash n%
* Thought:\textbackslash n%
Identify frequently interacted categories, genres, and item types.
Consider recency when it is relevant and identify patterns among
items the user has consistently rated highly. Compare each
candidate with the preferences inferred from these interaction
signals. Assign a preference score in the range 0 to 1 and give
a brief interaction-based rationale.\textbackslash n\textbackslash n%
When a corresponding Semantic Agent judgment is available, mark
the critique as ACCEPT or REJECT and include a brief reason in
the rationale. Use N/A when no corresponding semantic judgment
is available, including in the first round.\textbackslash n\textbackslash n%
Provide exactly one record for every eligible candidate in the
following format:\textbackslash n\textbackslash n%
CANDIDATE\_SCORES:\textbackslash n%
\textless{}item\_id\textgreater{} -- interaction-based rationale --
score (0\textasciitilde{}1) --
critique: \textless{}ACCEPT/REJECT/N/A\textgreater{}
\textbackslash n\textbackslash n%
* Action:\textbackslash n%
Select the Top-10 candidates with the highest preference scores
in CANDIDATE\_SCORES and list them in descending score order:
\textbackslash n\textbackslash n%
MY\_RANKING:\textbackslash n%
Rank. \textless{}item\_id\textgreater{} -- interaction-based rationale
-- score (0\textasciitilde{}1)\textbackslash n\textbackslash n%
* Constraint:\textbackslash n%
Evaluate only candidate-pool items absent from the user's
interaction history. Base your preference scores and rationales
on the supplied interaction signals. In MY\_RANKING, use the
scores reported in CANDIDATE\_SCORES.\textbackslash n\textbackslash n%
* Debate Policy:\textbackslash n%
Revise an interaction-based judgment when sufficiently supported
semantic feedback challenges an assumption or helps identify a
relevant preference signal in the user's history. Ground each
revision in the interaction evidence and briefly justify any
retained disagreement.\textbackslash n\textbackslash n%
Goal:\textbackslash n%
Produce history-grounded candidate evaluations and a Top-10
ranking derived from the interaction-based preference scores."
\end{tcolorbox}

\begin{tcolorbox}[
    title={Prompt Template for the Semantic Agent},
    colback=gray!5,
    colframe=gray!60,
    breakable,
    fonttitle=\bfseries
]
\small
\ttfamily
"You are the Semantic Preference Expert.\textbackslash n\textbackslash n%
Scope:\textbackslash n%
Produce an independent semantic assessment of how each candidate
matches the user's preferences. Ground factual claims in the
item-level evidence supplied by EG-RAR. Challenge the Interaction
Agent's judgments when the evidence supports a materially
different assessment.\textbackslash n\textbackslash n%
Task:\textbackslash n%
Evaluate every eligible item in the candidate pool. Consider
themes, narrative styles, item attributes, and semantic similarity
when assessing each item's relevance to the user's preferences.
For each eligible item, produce a Claim-Evidence record containing
a factual claim, its corresponding evidence passage, a preference
score, a confidence score, and a critique of the Interaction
Agent's judgment.\textbackslash n\textbackslash n%
Instructions:\textbackslash n\textbackslash n%
Follow three steps: Observation, Thought, and Output.
\textbackslash n\textbackslash n%
* Observation:\textbackslash n%
Review the Interaction Agent's current reasoning trace, including
its candidate-level evaluations and ranking. Identify agreements,
disagreements, and judgments that warrant further examination.
\textbackslash n\textbackslash n%
* Thought:\textbackslash n%
Infer the user's semantic preferences from their interaction
history and profile. Pay particular attention to the themes,
styles, narrative patterns, and attributes of items the user rated
highly. Examine whether interaction frequency, recency, or ratings
lead the Interaction Agent to overlook relevant item
characteristics.\textbackslash n\textbackslash n%
For each eligible candidate, use the structured semantic
representation to assess how well the item matches the user's
preferences. State one concise factual claim about the item that
informs this assessment. Quote the corresponding evidence passage
verbatim from the original information of the same item.
\textbackslash n\textbackslash n%
Assign a preference score in the range 0 to 1 to reflect the item's
semantic alignment with the user's preferences. Assign a confidence
score in the same range to reflect how certain you are about this
assessment. Express uncertainty and assign a lower confidence score
when the supplied evidence is limited. Indicate whether you accept
or reject the Interaction Agent's judgment and give a brief reason.
\textbackslash n\textbackslash n%
* Output:\textbackslash n%
Briefly assess the Interaction Agent's overall ranking and explain
your assessment. Then provide exactly one record for every eligible
candidate in the following format:\textbackslash n\textbackslash n%
CLAIM\_EVIDENCE\_RECORDS:\textbackslash n%
\textless{}item\_id\textgreater{} --
Claim: \textless{}factual claim about the item\textgreater{} --
Evidence: \textless{}verbatim passage from the item's original information\textgreater{} --
Preference score: \textless{}0\textasciitilde{}1\textgreater{} --
Confidence: \textless{}0\textasciitilde{}1\textgreater{} --
Critique: \textless{}ACCEPT/REJECT followed by a brief reason\textgreater{}
\textbackslash n\textbackslash n%
* Constraint:\textbackslash n%
Limit your evaluation to items in the candidate pool that are
absent from the user's interaction history. Use only the supplied
item information as the factual basis for each claim. Each quoted
passage must come from the item identified in the corresponding
record.\textbackslash n\textbackslash n%
* Debate Policy:\textbackslash n%
Maintain an independent semantic assessment. Accept the Interaction
Agent's judgment when it is compatible with the supplied evidence
or when the evidence provides insufficient grounds for challenging
it. Preserve and justify disagreement when the evidence reveals a
clear inconsistency or supports a materially different semantic
assessment.\textbackslash n\textbackslash n%
Goal:\textbackslash n%
Identify semantic preference patterns and produce an
evidence-grounded Claim-Evidence record and critique for every
eligible candidate."
\end{tcolorbox}

\begin{tcolorbox}[
    title={Prompt Template for the Judge Agent},
    colback=gray!5,
    colframe=gray!60,
    breakable,
    fonttitle=\bfseries
]
\small
\ttfamily
"You are the Judge Agent, responsible for decision-level ranking
calibration in MADRec.\textbackslash n\textbackslash n%
Scope:\textbackslash n%
Integrate the Interaction Agent's and Semantic Agent's outputs
after multi-round debate. Consider the user history and profile,
the accumulated debate history, both final Top-10 rankings, and
the complete candidate-level records. The semantic records include
the ECV support coefficients and adjusted semantic scores supplied
by EG-RAR.\textbackslash n\textbackslash n%
Task:\textbackslash n%
Construct the judge candidate set from the union of the two
agents' final Top-10 rankings. Estimate the relative reliability
of the interaction-based and semantic perspectives. For each item
in the union, combine the stored final-round scores using the
estimated weights. Select the Top-10 items with the highest
calibrated scores.\textbackslash n\textbackslash n%
Follow two steps: Thought and Action.\textbackslash n\textbackslash n%
1. Thought:\textbackslash n%
Review where the agents agree and how their disagreements
persist or are resolved across rounds. Examine whether revisions
are supported by the corresponding interaction signals or item-level
evidence. Consider which candidates were selected by both agents
and which were selected by one perspective. Use the supplied ECV
support coefficients as cues about the factual grounding of the
semantic judgments. Estimate the two weights from these cues,
the user history and profile, and the debate outcomes.
\textbackslash n\textbackslash n%
The reliability weights must satisfy:\textbackslash n%
omega\_I + omega\_S = 1\textbackslash n%
omega\_I, omega\_S in [0,1]\textbackslash n\textbackslash n%
WEIGHTS:\textbackslash n%
omega\_I = \textless{}value\textgreater{}, omega\_S =
\textless{}value\textgreater{}\textbackslash n\textbackslash n%
2. Action:\textbackslash n%
Construct the judge candidate set as the union of the two final
Top-10 rankings. For every item in this union, retrieve the
Interaction Agent's stored final-round preference score and the
adjusted semantic score computed by EG-RAR. Retrieve both scores
from the complete final-round candidate records, including for
items that appear in only one Top-10 ranking.
\textbackslash n\textbackslash n%
The supplied adjusted semantic score already incorporates the
ECV support coefficient. Use it directly in the following
calculation:\textbackslash n\textbackslash n%
Calibrated score = omega\_I * interaction-based preference score
+ omega\_S * adjusted semantic score\textbackslash n\textbackslash n%
Output the candidate set:\textbackslash n\textbackslash n%
JUDGE\_CANDIDATE\_SET:\textbackslash n%
\textless{}item\_id\textgreater{}, \textless{}item\_id\textgreater{},
...\textbackslash n\textbackslash n%
Provide exactly one calibrated record for every item in the judge
candidate set:\textbackslash n\textbackslash n%
CALIBRATED\_CANDIDATE\_SCORES:\textbackslash n%
\textless{}item\_id\textgreater{} --
interaction contribution:
\textless{}omega\_I * interaction-based preference score\textgreater{} --
semantic contribution:
\textless{}omega\_S * adjusted semantic score\textgreater{} --
calibration rationale -- calibrated score:
\textless{}0\textasciitilde{}1\textgreater{}\textbackslash n\textbackslash n%
Select the Top-10 items in descending order of calibrated score
and output:\textbackslash n\textbackslash n%
FINAL\_TOP10:\textbackslash n%
Rank. \textless{}item\_id\textgreater{} -- calibration rationale
-- calibrated score (0\textasciitilde{}1)\textbackslash n\textbackslash n%
Constraint:\textbackslash n%
Use only items in the judge candidate set that are absent from
the user's interaction history. Calculate both contributions
from the supplied final-round scores and the estimated weights.
The calibrated score must equal the sum of the two contributions.
Use the calibrated scores to determine FINAL\_TOP10. Ground the
calibration rationales in the agents' outputs and supplied
context.\textbackslash n\textbackslash n%
Goal:\textbackslash n%
Produce a final Top-10 ranking that combines interaction-based
preference and evidence-adjusted semantic relevance according
to their reliability for the current recommendation context."
\end{tcolorbox}
\begin{tcolorbox}[
    title={Semantic Schema Prompt for Evidence Extraction},
    colback=gray!5,
    colframe=gray!60,
    breakable,
    fonttitle=\bfseries
]
\small
\ttfamily
""You are an evidence extraction assistant for a recommender system.\textbackslash n\textbackslash n
Input:\textbackslash n
You will receive static item-level information, including structured metadata and item-side descriptive content, such as title, year, genre or category, synopsis or product description, semantic keywords, and domain-specific item attributes. User-generated reviews, ratings, and popularity-related signals are excluded.\textbackslash n\textbackslash n
Task:\textbackslash n
Extract structured item-level evidence from the provided information. All extracted evidence must be explicitly grounded in the input text. Do not introduce unsupported facts or external knowledge.\textbackslash n\textbackslash n
Please extract the following three types of evidence:\textbackslash n\textbackslash n
1. Semantic characteristics (S):\textbackslash n
Identify the main themes, topics, genres or categories, styles, concepts, and other semantic attributes that characterize the item.\textbackslash n\textbackslash n
2. Functional characteristics (F):\textbackslash n
Identify the primary functions, purposes, features, benefits, or use cases of the item. For items without explicit functional information, extract the main experiential or content-related characteristics supported by the input.\textbackslash n\textbackslash n
3. Contextual characteristics (C):\textbackslash n
Identify the relevant usage scenarios, intended audiences, settings, temporal background, cultural context, or other contextual information associated with the item.\textbackslash n\textbackslash n
Output format:\textbackslash n
Return the result in the following structured format:\textbackslash n\textbackslash n
S: <concise semantic evidence>\textbackslash n
F: <concise functional evidence>\textbackslash n
C: <concise contextual evidence>\textbackslash n\textbackslash n
Constraints:\textbackslash n
Only use information explicitly supported by the provided item-level content.\textbackslash n
Do not infer facts from general knowledge or introduce information that is absent from the input.\textbackslash n
If a type of evidence is not clearly supported, write 'Not explicitly mentioned'.\textbackslash n
Keep each evidence field concise, item-specific, and traceable to the corresponding input information.\textbackslash n\textbackslash n
Goal:\textbackslash n
Produce reliable and structured item-level evidence for downstream evidence-grounded recommendation reasoning."
\end{tcolorbox}
\begin{table}[t]
\centering
\caption{Performance comparison under the non-sequential collaborative filtering setting. The best results are shown in bold, and the strongest non-MADRec baseline results are underlined.}
\label{tab:item_rec_results}
\resizebox{\textwidth}{!}{
\begin{tabular}{lcccccc}
\toprule
\multirow{2}{*}{Method}
& \multicolumn{2}{c}{ML-1M}
& \multicolumn{2}{c}{Amazon-Book}
& \multicolumn{2}{c}{Amazon-Beauty} \\
\cmidrule(lr){2-3}
\cmidrule(lr){4-5}
\cmidrule(lr){6-7}
& Recall@10 & NDCG@10
& Recall@10 & NDCG@10
& Recall@10 & NDCG@10 \\
\midrule
NeuMF
& $0.154{\pm0.007}$ &$0.0721{\pm0.002}$
& $0.051{\pm0.008}$ &$0.0204{\pm0.002}$
& $0.060{\pm0.010}$ &$0.0507{\pm0.014}$\\

NAIS
& $0.159{\pm0.009}$ &$0.0732{\pm0.004}$
& $0.062{\pm0.007}$ &$0.0225{\pm0.003}$
& $0.065{\pm0.012}$ &$0.0524{\pm0.012}$ \\

XSimGCL
& $0.162{\pm0.008}$ &$0.0744{\pm0.004}$
& $0.067{\pm0.009}$ &$0.0241{\pm0.002}$
& $0.063{\pm0.011}$ &$0.0519{\pm0.011}$ \\

AutoCF
& $\underline{0.170}{\pm0.012}$ &$0.0746{\pm0.003}$
& $0.071{\pm0.008}$ &$0.0258{\pm0.006}$
& $0.071{\pm0.015}$ &$0.0538{\pm0.009}$ \\

DCCF
& $0.169{\pm0.011}$ &$\underline{0.0751}{\pm0.007}$
& $\underline{0.079}{\pm0.012}$ &$\underline{0.0264}{\pm0.007}$
& $\underline{0.075}{\pm0.013}$ &$\underline{0.0549}{\pm0.013}$ \\

MADRec$_{N}$
& $\mathbf{0.196}{\pm0.010}$
& $\mathbf{0.0805}{\pm0.004}$
& $\mathbf{0.107}{\pm0.007}$
& $\mathbf{0.0326}{\pm0.005}$
& $\mathbf{0.112}{\pm0.012}$
& $\mathbf{0.0603}{\pm0.011}$ \\
\bottomrule
\end{tabular}
}
\end{table}
\section{Generalization with a Collaborative Filtering Backbone} 
\label{app:item_rec} 
To examine whether MADRec depends on sequential interaction modeling, we replace its sequential retrieval backbone with NAIS~\cite{he2018nais}, an item-based collaborative filtering model. NAIS estimates candidate relevance from the collaborative relations between each candidate and the user's historically interacted items, without modeling sequential transitions. This experiment evaluates whether MADRec remains effective when candidate retrieval and interaction reasoning are based on item--item collaborative signals rather than sequential dependencies.

\subsection{Dataset}
We use the same ML-1M, Amazon-Book, and Amazon-Beauty datasets as in the main experiments. The original data filtering procedure, item-side evidence corpus, and chronological leave-one-out evaluation protocol remain unchanged. For each user, the most recent interaction is used for testing, the second most recent interaction is used for validation, and the remaining interactions are used for training.

\subsection{Evaluation and Implementation Details} 
For candidate retrieval, we instantiate the retrieval component using NAIS~\cite{he2018nais}, denoted as MADRec$_N$. NAIS is trained independently and remains frozen during the subsequent agent-based ranking stage. Given a user's historical interactions as an unordered item set, NAIS computes relevance scores over the full item space based on candidate-dependent item--item collaborative relations. Unless otherwise specified, all other implementation settings, including the core LLM, prompting strategy, number of debate rounds, and evaluation, remain identical to those used in the main experiments.

\subsection{Baselines} 
We compare MADRec with representative neural and deep collaborative filtering models, including NeuMF~\cite{he2017neural}, NAIS~\cite{he2018nais}, XSimGCL~\cite{yu2023xsimgcl}, AutoCF~\cite{autocf}, and DCCF~\cite{DCCF}. All baselines are evaluated using their original configurations.

\subsection{Results and Analysis}

Table~\ref{tab:item_rec_results} reports the results under the
non-sequential item-based recommendation setting.

As shown in Table~\ref{tab:item_rec_results}, MADRec$_N$ achieves the best overall performance across the evaluated datasets. It consistently outperforms its NAIS retrieval backbone, demonstrating that the performance gains arise not only from item-based candidate retrieval, but also from the subsequent multi-agent reasoning and ranking calibration. By jointly considering collaborative signals from historical interactions and evidence-grounded semantic information, MADRec$_N$ can more effectively distinguish relevant candidates within the retrieved candidate set.

MADRec$_N$ also performs favorably against representative neural and deep collaborative filtering models, including NeuMF, XSimGCL, AutoCF, and DCCF. Although these methods learn expressive user--item representations from collaborative signals, they primarily rely on implicit latent representations for recommendation. In contrast, MADRec$_N$ introduces explicit reasoning over candidate-level collaborative and semantic evidence. The \textbf{MAD} allows the \textit{Interaction Agent} and \textit{Semantic Agent} to identify disagreements between different signals. \textbf{EG-RAR} uses item-level evidence to evaluate semantic claims and adjusts their scores based on the support probabilities predicted by ECV, while \textbf{JRC} performs reliability-aware calibration of the interaction-based and semantic ranking outputs to produce the final ranking.

Together with the sequential recommendation results reported in the main paper, these findings demonstrate that MADRec is not restricted to a specific sequential retrieval architecture. Its \textbf{MAD}, \textbf{EG-RAR}, and \textbf{JRC} modules remain effective when the retrieval backbone is replaced by an item-based collaborative filtering model, supporting the generalizability of MADRec across different recommendation paradigms.

\section{ECV Data Construction and Training}
\label{app:ecv}

\subsection{Data Construction}

For each recommendation dataset, we construct an ECV dataset
from its item-level evidence base. Before generating any
claims, we randomly partition the items into disjoint
training, validation, and test groups, targeting 80\%, 10\%,
and 10\% of the items, respectively. Items that share an
identical source description or evidence passage are assigned
to the same group.

The Semantic Agent independently generates claims for items
sampled from each group. For each item $i$, the agent receives
its original information $x_i$ and structured semantic
representation $\phi_i$. It follows the same factual-claim
instructions used during recommendation inference to generate
concise statements about the item's characteristics. We pair
each statement with its corresponding evidence passage from
$x_i$. The pairs derived from the training, validation, and
test items form the ECV training, validation, and test sets,
respectively.

We remove duplicate evidence-claim pairs from each ECV
training set. The next subsection describes how we label
the pairs in each split. We combine the labeled ECV
training sets from the three recommendation datasets to
form the verifier training set $\mathcal{D}_{\mathrm{ver}}$.
\subsection{Annotation Protocol}
\begin{table}[t]
\centering
\caption{ECV dataset statistics and test-set annotation
agreement. S and NS denote the numbers of
\textnormal{Supported} and \textnormal{NotSupported} pairs,
respectively. The test counts are based on the final labels
after adjudication. Cohen's~$\kappa$ is computed from the
two independent test annotations before adjudication.}
\label{tab:ecv_data_stats}
\small
\begin{tabular}{@{}lcccc@{}}
\toprule
Dataset & Train (S / NS) & Valid (S / NS) & Test (S / NS)
& Test Cohen's~$\kappa$ \\
\midrule
ML-1M         & 3,196 / 836   & 403 / 109 & 400 / 100 & 0.83 \\
Amazon-Book   & 3,420 / 1,080 & 429 / 136 & 380 / 120 & 0.80 \\
Amazon-Beauty & 3,078 / 686   & 389 / 87  & 410 / 90  & 0.84 \\
\bottomrule
\end{tabular}
\end{table}
Each claim is assessed solely against its paired evidence
passage. A pair is labeled $\mathrm{Supported}$ if the
passage supports all factual assertions in the claim. It is
labeled $\mathrm{NotSupported}$ if any assertion lacks
sufficient support or conflicts with the passage. 

GPT-4o labels the training and validation sets. Two human
annotators independently label each test pair using the same
criteria. They receive only the evidence passage and its
paired claim. We resolve disagreements through adjudication
and use the final labels as the reference for the ECV
verification experiment. Table~\ref{tab:ecv_data_stats}
reports the number of pairs in each class for every dataset
and split, using the final labels for the test counts. We
also report Cohen's~$\kappa$ for each test set, computed from
the two independent annotations before adjudication.
\subsection{Training Procedure}

ECV uses a DeBERTaV3-base encoder and a binary classification
head. The encoder jointly processes an evidence passage and
its paired factual claim. We combine the labeled ECV training sets from the three
recommendation datasets to form $\mathcal{D}_{\mathrm{ver}}$.
We fine-tune both components on this set by minimizing the
cross-entropy loss in Eq.~\ref{eq:verifier_loss}.

The combined input is limited to 512 tokens, including
special tokens. If an input exceeds this limit, we retain
the full claim and truncate the evidence passage to fit the
remaining token budget. The same input preparation is used
during training and inference.

We use AdamW with a learning rate of $2\times10^{-5}$, a batch size of 16, and a weight decay of 0.01. We train the model for three epochs with a fixed random seed of 42 and select the checkpoint with the highest Macro-F1 on the pooled validation set. We evaluate this checkpoint on each held-out ECV test set and report the verification metrics.

For the downstream recommendation experiments, we use the
validation-selected checkpoint from the run with the
prespecified seed of 42. The verifier remains frozen
throughout recommendation inference.